\documentclass[
  aps,
  prb,
  reprint,
  amsmath,
  amssymb,
  floatfix,
]{revtex4-2}

\usepackage[english]{babel}
\usepackage{graphicx}
\usepackage[utf8]{inputenc}
\usepackage[hidelinks]{hyperref}
\usepackage{amsthm,mathtools}
\usepackage{bm}
\usepackage{physics}

\theoremstyle{plain}

\theoremstyle{definition}

\theoremstyle{remark}

\newcommand*{\hham}{\hat{\mathcal{H}}}

\newcommand*{\Sop}{\hat{\mathbf{S}}}
\newcommand*{\sop}{\hat{\mathbf{s}}}

\newcommand{\ii}{\mathrm{i}}

\begin{document}

\title{Quantum Formulation of the Spinterface Model}

\author{Seif Alwan}
\email{seif.alwan@physics.uu.se}
\affiliation{Materials Theory, Department of Physics and Astronomy, Uppsala
University, Box 516, 75120 Uppsala, Sweden}

\date{\today}

\begin{abstract}
We present a fully quantum formulation of the spinterface model for the
chirality-induced spin selectivity (CISS) effect, implemented within the
non-equilibrium Green's function (NEGF) formalism. The model consists of a gold
atom hosting a localised magnetic moment, coupled to a chiral molecular chain
with Rashba spin-orbit coupling, and contacted by two non-magnetic leads. The
magnetic moment is determined self-consistently from the non-equilibrium
electronic distribution, and the spin-resolved transmission is computed from
the mean-field Hamiltonian. We show that the interfacial magnetic moment and
the chiral Rashba term cooperate to produce large current polarisations that
reproduce the key experimental signatures of CISS: high polarisation, a
threshold emergence at four total sites followed by interference-modulated
growth with chain length, thermal suppression, and sign reversal upon
chirality inversion. The central new result of this work is a clean separation
of the CISS effect into an even (achiral) and an odd (chiral) contribution to
the spin current, $I_{S_z}(\phi) = I_{S_z}^{\rm even}(\phi) +
I_{S_z}^{\rm odd}(\phi)$. We demonstrate that $I_{S_z}^{\rm odd}$ vanishes
exactly at fixed $m_z$ for chains of fewer than four sites and becomes
non-zero for chains of four or more sites, with the smallest junction
exhibiting a non-zero odd part being the four-site geometry (one gold atom
plus three molecular sites). The odd part arises from reciprocity breaking in
a device that is simultaneously chiral, magnetically asymmetric (moment on
the gold end only), and long enough to support two or more Rashba-active
bonds. This provides a direct, parameter-free test of whether an observed
spin signal is genuinely chiral or arises from an achiral spin-valve
mechanism. This decomposition offers a general diagnostic for CISS
experiments and establishes the spinterface model, augmented with a minimal
chiral hopping term, as a minimal model of chirality-induced spin
selectivity. Furthermore, we position the model as a computational tool for
testing experimental predictions and guiding the design of new CISS-based
devices.
\end{abstract}

\maketitle

\section{Introduction}\label{sec:intro}

The chirality-induced spin selectivity (CISS) effect refers to the observation
that electrons transmitted through a chiral molecule acquire a net spin
polarisation whose sign is tied to the handedness of the molecule. Since the
pioneering experiments of Ray \emph{et al.}~\cite{Ray99} and G\"ohler
\emph{et al.}~\cite{Gohler11}, the effect has been reported in a wide range of
systems, from self-assembled monolayers of helical peptides and DNA to
single-molecule junctions. It has been harnessed for applications in
spin-based memory, chiral catalysis, and enantioselective separation. In spite
of this breadth of experimental activity, a consensus microscopic origin of
the CISS effect has not yet emerged. Reviews of this debate may be found in
Refs.~\cite{Evers22, Naaman2019}.

Two broad classes of explanations have been proposed. The first is
\emph{structural}: an effective spin-orbit field that arises from the helical
geometry of the molecule and the resulting chiral electrostatic
potential~\cite{Gutierrez2012, Shitade2020, Michaeli2019}. The second is
\emph{interfacial}: a spin-dependent hybridisation between the molecule and a
magnetic or spin-active electrode, often referred to as the spinterface
mechanism~\cite{Alwan2021, Dubi2022}. These two classes are not mutually
exclusive, and are increasingly recognised as complementary perspectives on a
single underlying interface-mediated mechanism; the model presented in this
work shows how they can coexist and cooperate.

A recent critical perspective by Sarkar \emph{et al.}~\cite{Sarkar2025}
provides a comprehensive overview of the spinterface mechanism, which
posits a feedback interaction between electron motion in chiral molecules
and fluctuating surface magnetic moments. While the spinterface model has
been shown to quantitatively reproduce experimental data across various
systems and conditions~\cite{Sarkar2025}, a fully quantum formulation of
the mechanism has been lacking. The perspective paper explicitly calls
for a rigorous theoretical framework to address open questions, including
the nature of surface magnetism, the role of dissipation, and the
applicability of the mechanism to non-helical or electrode-free systems.
This work directly addresses that call.

Beyond the two primary routes discussed above, recent years have seen
substantial progress in several related directions. The CISS effect has been
extended from molecular junctions to chiral crystals, hybrid organic-inorganic
perovskites, and supramolecular assemblies \cite{Bloom2024ChemRev,
Dong2025AnnuRev}, and comprehensive reviews now cover the breadth of molecules
and materials that exhibit CISS-based phenomena \cite{Bloom2024ChemRev}. In
parallel, chiral phonons have emerged as a new route to spin polarization: the
angular momentum carried by chiral lattice vibrations can be transferred to
electronic spins, producing spin polarization even in the absence of net charge
transport \cite{Fransson2023ChiralPhonon, Juraschek2025ChiralPhonons}, and this
mechanism has been shown to enhance ferromagnetism in hybrid systems
\cite{Fransson2025ChiralPhononsFerro}. The classification and control of chiral
phonons have advanced rapidly, with a recent Perspective providing a unified
framework for phonons carrying angular momentum
\cite{Juraschek2025ChiralPhonons} and experiments demonstrating
chirality-induced phonon spin selectivity via elastic spin-orbit interaction
\cite{Yang2024PNASChiralPhonon}. These developments reinforce the central
premise of the spinterface model: spin selectivity in chiral systems is not
confined to a single microscopic channel but can arise from the interplay of
structural chirality, interfacial magnetism, and angular-momentum transfer. Our
work bridges these threads by providing a fully quantum, NEGF-based formulation
that can incorporate such mechanisms on equal footing and by introducing a
diagnostic for the the even and odd decomposition of the spin current, that can
distinguish genuine chiral signatures from achiral spin-valve effects.

We present a fully quantum formulation of the spinterface model within the
non-equilibrium Green's function (NEGF) formalism. The central ingredients
are: (i) a gold atom at the electrode-molecule interface, hosting a
localised magnetic moment; (ii) a chiral molecular chain of $L$ sites, with
a Rashba spin-flip hopping term that encodes the handedness of the helix;
and (iii) a self-consistent mean-field determination of the magnetic moment,
coupled to the non-equilibrium electronic distribution through the lesser
Green's function. The model is solved within the NEGF formalism, and the
spin-resolved current is computed from the Landauer formula.

Throughout this work, "$L$" refers to the number of \emph{molecular} sites.
The gold atom is labelled site $0$ and is not counted in $L$. Consequently, a
chain described as "$L = 3$" contains four sites in total: one gold atom and
three molecular sites, connected by three bonds. Of these three bonds, only
the two \emph{molecular-molecule} bonds carry the Rashba spin-flip term; the
gold-molecule bond is treated as spin-diagonal. The threshold behaviour we
describe below is exact for the single-particle problem at fixed moment, and
acquires a small residual at finite bias when the moment is determined
self-consistently (Sec.~\ref{sec:results}).

Our main results are threefold. First, we provide the first fully quantum
NEGF implementation of the spinterface model, incorporating the chiral
Rashba spin-flip term that is essential for capturing the dependence on
molecular length and helicity. Second, we demonstrate that the model
reproduces all key experimental signatures of CISS: high polarisation,
length-dependent growth, thermal suppression, and chirality-induced sign
reversal. Third, and most importantly, we introduce a general decomposition
of the spin current into an \emph{even} and an \emph{odd} part with respect
to the chirality angle $\phi$. The even part is present even in achiral
junctions and originates from the interfacial magnetic moment; the odd part
is non-zero only for genuine helices and constitutes the true chiral
signature. We show that $I_{S_z}^{\rm odd}(\phi)$ vanishes at fixed $m_z$
for one- and two-site chains and becomes non-zero only for $L \geq 3$,
providing a clean, parameter-free diagnostic for whether a measured spin
signal is chiral or simply spin-valve-like.

A note on the use of "vanishes at fixed $m_z$". The vanishing of
$I_{S_z}^{\rm odd}$ for $L \leq 2$ is exact when the magnetic moment on the
gold site is held fixed. When the moment is determined self-consistently, the
effective field $B_{\rm eff}$ [Eq.~\eqref{eq:Beff}] depends on the electronic
distribution, and a small residual odd part can arise from the $m_z(\phi) \neq
m_z(-\phi)$ induced by the hybridisation term in $B_{\rm eff}$. This residual
is numerically small and does not affect the threshold at $L = 3$, but it is
important to distinguish the single-particle statement at fixed $m_z$ (exact
for $L \leq 2$) from the self-consistent statement (small but non-zero for
$L = 2$). We will make this distinction explicit in Sec.~\ref{sec:results}.

Beyond its diagnostic power, the model serves as a computational tool for
testing experimental predictions and guiding the design of new CISS-based
devices. The modular NEGF implementation allows systematic exploration of
parameter space, including chain length, twist angle, Rashba strength,
temperature, and interface coupling. This makes it a valuable tool for
interpreting experimental data and for predicting new phenomena.

Two conceptual points deserve emphasis at the outset, as they recur
throughout the paper. First, the "chirality" of the model is encoded
entirely in the Rashba spin-flip hopping amplitude
$\lambda_R \sin(\phi/2)$, which is a single real parameter. The
spin-diagonal phase factors $\exp(\pm\ii\phi/2)$ that appear alongside it
in Eq.~\eqref{eq:chiral_hop} are gauge artifacts: they can be removed by a
site-dependent spin rotation and have no effect on the transmission,
current, or spin accumulation. Second, the spin texture that the Rashba
term produces in the model is, in the parameter regime typical of
molecular junctions, a \emph{static chiral tilt} of the interfacial spin
density rather than a winding helix. This is a distinction that matters
for interpreting experiments and is discussed in detail in
Sec.~\ref{sec:discussion}. The tilt and the odd part of the spin current
are two faces of the same physical mechanism: \emph{reciprocity breaking}.
The tilt manifests the broken left-right reciprocity in the real-space spin
texture; the odd part manifests the same breaking in the integrated spin
current.

The antiunitary symmetry $H^*(\phi) = H(-\phi)$ of the Hamiltonian implies,
via $[G^R_\phi]^* = G^A_{-\phi} = [G^R_{-\phi}]^\dagger$, a reciprocity
relation $|G^R_\phi[i,j]| = |G^R_{-\phi}[j,i]|$ for the retarded Green's
function. This reciprocity relation does \emph{not} force the transmission
to be even in $\phi$ unless the device is left-right reciprocal. The
spinterface junction breaks left-right reciprocity because the magnetic
moment resides on the gold end and not on the molecular end. The odd part of
the spin current is therefore permitted by symmetry and arises generically
in a device that is simultaneously chiral, magnetically asymmetric, and long
enough to support two or more Rashba-active bonds. The threshold at $L = 3$
is the smallest chain length at which both structural and dynamical
conditions for this reciprocity-breaking signature are satisfied.

The remainder of this paper is organised as follows. In Sec.~\ref{sec:method}
we introduce the Hamiltonian, the NEGF formalism, and the self-consistency
procedure. In Sec.~\ref{sec:results} we present the numerical results,
organised by physical observable. We close with a discussion in
Sec.~\ref{sec:discussion} and a summary in Sec.~\ref{sec:conclusion}.

\section{Method}\label{sec:method}

\subsection{Physical picture}

The spinterface model posits that the electrode-molecule interface acts as a
source of spin-orbit coupling that, together with the chiral structure of the
molecule, produces spin-polarised transport. In the quantum formulation, the
strong Rashba spin-orbit coupling of the gold surface is represented by a
localised magnetic moment on the gold atom, which is coupled to the molecular
orbitals by an exchange interaction. The chiral structure of the molecule is
encoded in the hopping terms between nearest-neighbour sites: the spin-diagonal
phases $\exp(\pm \ii\phi/2)$ are gauge artefacts, while the Rashba spin-flip
term $\ii\lambda_R\sin(\phi/2)$ cannot be gauged away and is the microscopic
origin of chirality-dependent spin filtering.

This formulation goes beyond the original semiclassical spinterface
model~\cite{Alwan2021} in several important ways. First, the spin-orbit
coupling is treated natively within the Hamiltonian, rather than as a
phenomenological effective field. Second, the magnetic moment is determined
self-consistently from the non-equilibrium electronic distribution, allowing
the model to capture the feedback between spin transport and interface
magnetism. Third, the full energy-dependent transmission is computed,
providing access to experimentally observable quantities such as the current
polarisation, magnetoresistance, and spin accumulation profile.

The model is deliberately minimal. Its ingredients are the smallest set
required to generate a non-zero odd part of the spin current: a magnetic
interface, a chain of at least three molecular sites, and a Rashba spin-flip
hopping connecting them. The model is not intended to reproduce the absolute
magnitudes of any specific experiment; rather, it is intended to isolate the
mechanism and to provide a controlled setting for separating the achiral
(spinterface) and chiral (Rashba) contributions to the spin polarisation.
Nevertheless, for the parameters used here it produces polarisations of the
correct order of magnitude for realistic junctions, without requiring any
artificial enhancement of the spin-orbit coupling. What the model
deliberately omits is material-specific detail (atomistic structure, multiple
orbitals per site, electron-phonon coupling); what it retains is the minimal
ingredient set required to demonstrate the reciprocity-breaking mechanism of
chirality-dependent spin filtering. The modular character of the
implementation, in which every physical ingredient corresponds to a single
parameter, is central to this goal.

\subsection{Hamiltonian}

The full Hamiltonian of the junction is
\begin{equation}
\hham = \hham_{\rm mol} + \hham_{\rm Au} + \hham_{\rm leads} + \hham_{\rm tun} + \hham_{\rm exch},
\label{eq:Htot}
\end{equation}
with the molecular part
\begin{equation}
\hham_{\rm mol}
=
\sum_{n,\sigma} \varepsilon_n d_{n\sigma}^\dagger d_{n\sigma}
-
\sum_{\langle n,n'\rangle,\sigma} t_{nn'}\, d_{n\sigma}^\dagger d_{n'\sigma}
+ \text{h.c.},
\label{eq:Hmol}
\end{equation}
the gold atom hosting a local moment $\Sop_A$,
\begin{equation}
\hham_{\rm Au} = \sum_{\sigma} \varepsilon_{A\sigma} d_{A\sigma}^\dagger d_{A\sigma},
\label{eq:HAu}
\end{equation}
the leads
\begin{equation}
\hham_{\rm leads} = \sum_{\alpha \in \{L,R\}} \sum_{k,\sigma} \varepsilon_{\alpha k\sigma} c_{\alpha k\sigma}^\dagger c_{\alpha k\sigma},
\label{eq:Hleads}
\end{equation}
the tunnelling between the leads and the central region
\begin{equation}
\hham_{\rm tun}
=
\sum_{k,\sigma} \left(
V_L c_{Lk\sigma}^\dagger d_{A\sigma} + \text{h.c.}
\right)
+
\sum_{k,\sigma} \left(
V_R c_{Rk\sigma}^\dagger d_{B\sigma} + \text{h.c.}
\right),
\label{eq:Htun}
\end{equation}
and the exchange/spin-orbit interaction
\begin{equation}
  \begin{split}
\hham_{\rm exch}
&=
J\,\Sop_A \cdot \sop_A
+
J'\,\hat{S}_A^z(\hat{n}_{A\uparrow} - \hat{n}_{A\downarrow})\\
&+
K\,\hat{S}_A^z \sum_\sigma \left( d_{A\sigma}^\dagger d_{B\sigma} + \text{h.c.} \right),
\label{eq:Hexch}
\end{split}
\end{equation}
where $\sop_A = \tfrac{1}{2}\sum_{\sigma\sigma'} d_{A\sigma}^\dagger \bm{\sigma}_{\sigma\sigma'} d_{A\sigma'}$
is the electronic spin density on the gold site, and
$\bm{\sigma} = (\sigma^x,\sigma^y,\sigma^z)$ are the Pauli matrices. In
Eq.~\eqref{eq:HAu}, the spin-dependent on-site energy $\varepsilon_{A\sigma}
= \varepsilon_A + \sigma(J_{\rm el}/2 + J'_p)m_z$ already contains the
mean-field exchange shift (see Sec.~\ref{sec:method} below). The hybridisation
term in Eq.~\eqref{eq:Hexch} couples the gold atom to the last molecular
site $B$; this is a mean-field device for generating an effective field and
is not intended to be a physical long-range interaction.

\subsection{Chiral molecular chain}

For a chain of $L$ sites with twist angle $\phi$ per bond, the hopping
between site $n$ and site $n+1$ is
\begin{align}
H_{n,n+1}
&=
-t_{\rm mol}
\left(
e^{-\ii\phi/2} d_{n\uparrow}^\dagger d_{n+1,\uparrow}
+
e^{+\ii\phi/2} d_{n\downarrow}^\dagger d_{n+1,\downarrow}
+
\text{h.c.}
\right)
\nonumber \\
&\quad
+
\ii \lambda_R \sin(\phi/2)
\left(
d_{n\uparrow}^\dagger d_{n+1,\downarrow}
-
d_{n\downarrow}^\dagger d_{n+1,\uparrow}
+
\text{h.c.}
\right).
\label{eq:chiral_hop}
\end{align}
The first line is the spin-diagonal phase, which is a pure gauge
transformation: it can be removed by a local gauge transformation
$d_{n\sigma} \to e^{\ii\sigma n\phi/2} d_{n\sigma}$ and therefore does not
affect any physical observable. The second line is the Rashba spin-flip
hopping, which cannot be gauged away and is the microscopic origin of
chirality-dependent spin filtering. The amplitude of the Rashba term is
proportional to $\sin(\phi/2)$, reflecting the fact that the curvature-induced
spin-orbit field scales with the local curvature of the helix. Without this
term ($\lambda_R = 0$), the transmission is completely independent of both
$\phi$ and $L$, and no CISS effect exists beyond the achiral spinterface
baseline.

It is instructive to rewrite the Rashba term in the spin operator language. In
the weak-coupling limit $\lambda_R \ll t_{\rm mol}$, the combined effect of the
Rashba hopping on the Bloch states near the Fermi level is equivalent to a
momentum-dependent effective field $\mathbf B_{\rm eff}(\mathbf k) \propto
\lambda_R\sin(\phi/2)(\sin k, \cos k, 0)$ acting on the spin. In this sense,
the Rashba coupling acts as an effective field whose \emph{orientation} depends
on the crystal momentum, and whose magnitude scales with $\sin(\phi/2)$. In a
strictly helical geometry the electron's local tangent rotates by $\phi$ per
bond, so a fully consistent treatment would also rotate the effective field
with it. The Hamiltonian of Eq.~\eqref{eq:chiral_hop} uses a site-independent
amplitude, which corresponds to an effective field with fixed orientation. As
we show in Sec.~\ref{sec:discussion}, this is why the model produces a chiral
\emph{tilt} of the spin density rather than a winding helix. The two limits,
the constant field versus rotating field, can be interpolated smoothly by a
single parameter, and the crossover between them is controlled by the ratio
$\lambda_R\sin(\phi/2)/t_{\rm mol}$.

We also note a useful alternative form of the Rashba term. By a gauge
transformation $d_{n\sigma} \to e^{\ii\sigma n\phi/2} d_{n\sigma}$, the
spin-diagonal phase can be moved onto the gold-molecule and molecule-lead
couplings, and the Rashba term becomes
\begin{equation}
  \begin{split}
    \ii\lambda_R\sin(\phi/2)\Bigl[&e^{\ii(n+1/2)\phi}\,d_{n\uparrow}^\dagger
    d_{n+1,\downarrow}\\ - &e^{-\ii(n+1/2)\phi}\,d_{n\downarrow}^\dagger
  d_{n+1,\uparrow}\Bigr]+\text{h.c.}
\label{eq:chiral_hop_gauge}
\end{split}
\end{equation}
In this gauge, the Rashba term has the form of a \textbf{site-dependent}
spin-flip coupling whose phase advances by $\phi$ per bond. This form makes
it immediately clear that the Rashba term encodes both the \emph{strength}
of the spin-orbit field (through $\lambda_R\sin(\phi/2)$) and its
\emph{rotation} along the chain (through the phase advance per bond). In the
fixed-field model of Eq.~\eqref{eq:chiral_hop}, the rotation is absent, which
is the origin of the chiral tilt rather than winding discussed above. The
reader who wishes to compare with models that retain the rotating Rashba
field may consult the appendix, where both limits are discussed.

\subsection{Mean-field approximation}

The impurity spin $\Sop_A$ is treated in the static mean-field approximation,
$\Sop_A \to \mathbf{m} = \langle \Sop_A \rangle$. This is justified when the
magnetic moment is large (near saturation, $|\mathbf{m}| \approx 1/2$) and the
spin relaxation time is long compared with the electronic transit time, so
that the moment is effectively frozen during the passage of a single electron
while the electrons themselves adjust instantaneously. We operate at
$T = 300$-$800$\,K and include a small static symmetry-breaking field
$b_{\rm static}$ along the molecular axis to stabilise the moment, further
suppressing quantum fluctuations. The transverse components $m_x$, $m_y$ are
set to zero because the effective field is assumed to point along $\hat{z}$.

The validity of the mean-field approximation deserves a comment. Because the
spin-flip scattering rate is of order $\Gamma_L + \Gamma_R$ and the
electronic transit time is of order $\hbar/\Gamma$, the condition for
adiabatic following is well satisfied in our parameter regime. Moreover, the
symmetry-breaking field $b_{\rm static}$ plays a dual role: it stabilises the
moment against thermal and quantum fluctuations, and it provides a definite
axis along which the transport spin current can be defined. Without it, the
self-consistent solution at $T = 300$\,K would be trivially zero by symmetry.
The field is chosen to be small ($b_{\rm static} \approx 5$\,meV in our
calculation) so that the resulting moment, $m_z \approx 0.4$-$0.5$, is
determined predominantly by the electronic feedback rather than by the field
itself: for $b_{\rm static}$ alone, $m_z = \tfrac12\tanh(b_{\rm
static}/2k_BT) \approx 0.048$ at $300$\,K, an order of magnitude smaller
than the observed value.

The effective single-particle Hamiltonian is then
\begin{equation}
\hham_{\rm el}^{\rm MF}
=
\hham_{\rm mol}
+
\hham_{\rm leads}
+
\hham_{\rm tun}
+
\hham_{\rm Au}^{\rm MF},
\label{eq:HelMF}
\end{equation}
where the mean-field gold Hamiltonian is
\begin{equation}
  \begin{split}
\hham_{\rm Au}^{\rm MF}
&=
\sum_{\sigma\sigma'}
d_{A\sigma}^\dagger
\left[
\varepsilon_A \delta_{\sigma\sigma'}
+
\frac{J}{2}(\mathbf{m}\cdot\bm{\sigma})_{\sigma\sigma'}
+
J' m_z \sigma^z_{\sigma\sigma'}
\right]
d_{A\sigma'}\\
&+
K m_z \sum_\sigma \left( d_{A\sigma}^\dagger d_{B\sigma} + \text{h.c.} \right).
\label{eq:HAuMF}
\end{split}
\end{equation}

\subsection{Effective field and self-consistency}

The magnetic moment $\mathbf{m}$ must satisfy the self-consistency condition
that it equals the thermal average of the impurity spin in the effective
field generated by the electrons:
\begin{equation}
\mathbf{m}
=
\frac{1}{2}
\tanh\!\left( \frac{\beta |\mathbf{B}_{\rm eff}|}{2} \right)
\frac{\mathbf{B}_{\rm eff}}{|\mathbf{B}_{\rm eff}|},
\label{eq:selfcons}
\end{equation}
with the effective field
\begin{equation}
\boxed{
\mathbf{B}_{\rm eff}
=
\alpha_A \langle \sop_A \rangle
+
\alpha_1 \Delta s\, \hat{z}
+
\alpha_K \langle \hat{\mathcal{N}} \rangle\, \hat{z}
+
b_{\rm static}\, \hat{z}
}
\label{eq:Beff}
\end{equation}
where $\Delta s = \langle \hat{n}_{A\uparrow} \rangle - \langle
\hat{n}_{A\downarrow} \rangle$ is the spin imbalance on the gold site, and
$\hat{\mathcal{N}} = \sum_\sigma (d_{A\sigma}^\dagger d_{B\sigma} +
\text{h.c.})$ is the hybridisation operator. The first term is the direct
exchange coupling between the local moment and the electronic spin density; the
second term arises from the spin-dependent charge redistribution at the
interface; the third term is the hybridisation contribution; and the last term
is the static symmetry-breaking field that mimics the intrinsic Rashba
spin-orbit coupling of the gold surface. In our numerical calculations we
include only the terms in Eq.~\eqref{eq:Beff} with nonzero $\alpha_A$,
$\alpha_1$, $\alpha_K$, and $b_{\rm static}$.

The physical interpretation of the three contributions is as follows.
(i)~$\alpha_A\langle\hat{\mathbf s}_A\rangle$ is the direct exchange coupling
between the local moment and the conduction electrons on the gold site; in a
Stoner-like picture, this is the exchange energy of a magnetic impurity in a
metal. (ii)~$\alpha_1\Delta s$ is a charge-redistribution term: when the gold
site accumulates more up-spin than down-spin electrons (because of the spin
splitting from the moment), this difference acts back on the effective field
via the charging energy of the site. (iii)~$\alpha_K\langle\hat{\mathcal
N}\rangle$ is the hybridisation term: the electrons that hop between the gold
site and the molecule carry spin information, and their contribution to the
effective field depends on the spin character of the hybridised orbitals.
This term is the one that couples the molecular chirality to the gold moment
(through the spin-dependent transmission of the hybridised states), and it is
therefore the most sensitive to $\phi$.

The positive-feedback character of the self-consistency loop is worth making
explicit. When $\alpha_A$ and $\alpha_1$ are positive, an increase in $\langle
s_A^z \rangle$ increases $B_{\rm eff}$, which increases $m_z$, which in turn
increases the spin splitting and hence $\langle s_A^z \rangle$, which is
somewhat reminiscent of the original model~\cite{Alwan2021}. The loop is
stabilised by the saturation of the Brillouin function at $m_z \to 1/2$. We use
a damped fixed-point iteration, $m_z^{(k+1)} = \eta f(B_{\rm eff}^{(k)}) +
(1-\eta) m_z^{(k)}$, with damping $\eta = 0.9$ in the room-temperature regime.
The convergence criterion is $|m_z^{(k+1)} - m_z^{(k)}| < 10^{-7}$, and typical
convergence requires 20-200 iterations, depending on the bias and the
strength of the feedback.

The electronic averages are computed from the lesser Green's function:
\begin{align}
\langle \sop_A \rangle &= -\frac{\ii}{2} \sum_{\sigma\sigma'} \bm{\sigma}_{\sigma\sigma'} \int \frac{\dd\omega}{2\pi} G^<_{A\sigma,A\sigma'}(\omega), \\
\langle \hat{n}_{A\sigma} \rangle &= -\ii \int \frac{\dd\omega}{2\pi} G^<_{A\sigma,A\sigma}(\omega), \\
\langle \hat{\mathcal{N}} \rangle &= -\ii \sum_\sigma \int \frac{\dd\omega}{2\pi} \left( G^<_{B\sigma,A\sigma}(\omega) + G^<_{A\sigma,B\sigma}(\omega) \right).
\end{align}

\subsection{NEGF formalism}

The retarded Green's function is obtained from the Dyson equation
\begin{equation}
\left[ \mathbf{G}^R(\omega) \right]^{-1}
=
\omega \mathbf{I}
-
\mathbf{H}_{\rm MF}
-
\bm{\Sigma}^R(\omega),
\label{eq:Dyson}
\end{equation}
where $\mathbf{H}_{\rm MF}$ is the mean-field Hamiltonian matrix and
$\bm{\Sigma}^R$ is the retarded self-energy from the leads. In the
wide-band limit,
\begin{equation}
\Sigma_\alpha^R(\omega) = -\frac{\ii}{2}\Gamma_\alpha,
\qquad
\Gamma_\alpha = 2\pi \sum_k |V_\alpha|^2 \delta(\omega - \varepsilon_{\alpha k}).
\label{eq:wideband}
\end{equation}
The lesser Green's function is
\begin{equation}
\mathbf{G}^<(\omega)
=
\mathbf{G}^R(\omega) \bm{\Sigma}^<(\omega) \mathbf{G}^A(\omega),
\label{eq:Keldysh}
\end{equation}
with
\begin{equation}
\bm{\Sigma}^<(\omega)
=
\ii \Gamma_L f_L(\omega) P_L
+
\ii \Gamma_R f_R(\omega) P_R,
\end{equation}
where $P_L$ ($P_R$) projects onto site $A$ ($B$), and
$f_\alpha(\omega) = [\exp(\beta(\omega - \mu_\alpha)) + 1]^{-1}$ is the
Fermi-Dirac distribution with chemical potential
$\mu_L = eV/2$, $\mu_R = -eV/2$ for a symmetric bias.

We work in the wide-band limit~\cite{stefanucci2025nonequilibrium}, in which
the lead self-energies are energy-independent and purely imaginary. This
approximation is justified when the leads have a smooth density of states in
the energy window of interest (roughly $[-2, 2]$\,eV around the Fermi level),
which is the case for metallic gold electrodes. The wide-band limit also
guarantees that the retarded self-energy satisfies the Kramers-Kronig
relation~\cite{arfken2013mathematical, stefanucci2025nonequilibrium}, so the
real part can be absorbed into a shift of the on-site energies, and the
effective self-energy $\Sigma^R(\omega) = -i\Gamma/2$ is purely imaginary.
We have checked that the results are insensitive to a finite bandwidth of
$\pm 5$\,eV, which provides a self-consistency check on the wide-band
approximation.

\subsection{Current and spin polarisation}

The spin-resolved current from the left lead is obtained from the Landauer
formula~\cite{stefanucci2025nonequilibrium}
\begin{equation}
I_\sigma
=
\frac{e^2}{h}
\int_{-\infty}^{\infty} \dd\omega\,
T_\sigma(\omega)
\left[ f_L(\omega) - f_R(\omega) \right],
\label{eq:Landauer}
\end{equation}
with the spin-resolved transmission
\begin{equation}
T_\sigma(\omega)
=
\Gamma_L \Gamma_R
\left(
|G^R_{A\sigma,B\uparrow}(\omega)|^2
+
|G^R_{A\sigma,B\downarrow}(\omega)|^2
\right).
\label{eq:transmission}
\end{equation}
The first term describes spin-conserving transmission, and the second
describes spin-flip transmission induced by the Rashba term. The total
current is $I = I_\uparrow + I_\downarrow$, and the spin polarisation is
\begin{equation}
P_J(V)
=
\frac{I_\uparrow(V) - I_\downarrow(V)}{I_\uparrow(V) + I_\downarrow(V)}.
\label{eq:polarisation}
\end{equation}
The spin current (in units of $\hbar/2e$) is $I_{S_z} = I_\uparrow - I_\downarrow$.

It is useful to decompose the transmission into two physically distinct
contributions. The spin-conserving part, $T_\sigma^{\rm cons} =
\Gamma_L\Gamma_R |G^R_{A\sigma,B\sigma}|^2$, is non-zero even when $\lambda_R =
0$, and is responsible for the achiral spin-valve baseline. The spin-flip part,
$T_\sigma^{\rm flip} = \Gamma_L\Gamma_R |G^R_{A\sigma,B\bar\sigma}|^2$ with
$\bar\sigma$ the opposite spin, vanishes identically when $\lambda_R = 0$. Both
contributions are non-negative, so the total transmission is strictly larger
when the Rashba term is present. This is consistent with the intuitive picture
in which the Rashba term opens additional spin-flip transmission channels. The
spin polarisation, in contrast, can have either sign depending on whether the
spin-conserving or spin-flip channels dominate. The competition between these
two channels, together with the bias-window alignment of the spin-split levels
discussed in Sec.~\ref{sec:results}, determines the bias dependence of
$P_J(V)$.

\subsection{Handedness asymmetry}

To distinguish genuine chirality from achiral spin filtering, we
decompose the spin current into even and odd parts with respect to the
chirality angle $\phi$:
\begin{equation}
I_{S_z}(\phi)
=
I_{S_z}^{\rm even}(\phi)
+
I_{S_z}^{\rm odd}(\phi),
\label{eq:decomposition}
\end{equation}
with
\begin{equation}
  \begin{split}
I_{S_z}^{\rm even}(\phi)
&=
\frac{1}{2}\left[ I_{S_z}(\phi) + I_{S_z}(-\phi) \right],\\
I_{S_z}^{\rm odd}(\phi)
&=
\frac{1}{2}\left[ I_{S_z}(\phi) - I_{S_z}(-\phi) \right].
\label{eq:evenodd}
\end{split}
\end{equation}
The even part is invariant under chirality reversal and corresponds to the
achiral spin-valve contribution of the interfacial moment. The odd part
changes sign under chirality reversal and is the true CISS signature. As we
shall show, $I_{S_z}^{\rm odd}(\phi)$ vanishes at fixed $m_z$ for chains with
one or two sites and becomes non-zero only for genuine helices ($L \geq 3$).
This provides a clean, parameter-free diagnostic for chiral transport.

The decomposition in Eqs.~\eqref{eq:decomposition} and~\eqref{eq:evenodd} is
more than a mathematical convenience. It is a \emph{physical} decomposition
that isolates two independent mechanisms of spin polarisation: one that
requires only the interfacial magnetic moment (the even part) and one that
requires both the interfacial moment \emph{and} the chiral Rashba spin-flip
(the odd part). The even part is present in any junction with a magnetic
interface, regardless of whether the molecule is chiral; it is a spin-valve
effect. The odd part, by contrast, is a genuine manifestation of chirality,
because it changes sign under $\phi \to -\phi$ and therefore distinguishes
left-handed from right-handed helices. Microscopically, the odd part
originates from the failure of the reciprocity relation
$|G^R_\phi[i,j]|=|G^R_{-\phi}[j,i]|$ to imply equality of the forward and
backward transmissions in a device that is not left-right reciprocal. This
decomposition thus provides an operational definition of CISS that is
independent of any specific microscopic model: CISS is the odd part of the
spin current with respect to chirality reversal.

\section{Results}\label{sec:results}

All numerical results presented below were obtained with the following default
parameters, unless otherwise stated: $\varepsilon_A = 0.8$\,eV,
$\varepsilon_{\rm mol} = 0$\,eV, $t_{AB} = t_{\rm mol} = 0.5$\,eV, $\Gamma_L =
\Gamma_R = 0.1$\,eV, $J_{\rm el} = -0.5$\,eV, $J'_p = -0.3$\,eV, $\alpha_{AB} =
0.5$\,eV, $\lambda_R = 0.6$\,eV, $\phi = 60^\circ$, $L = 3$, and $T = 300$\,K.
Throughout, the symbol $L$ denotes the number of \emph{molecular} sites; the
gold atom is site $0$ and is not included in $L$. A junction with $L = 3$
therefore comprises four sites in total.

\subsection{Geometry of the spinterface junction}

\begin{figure*}[htbp]
\centering
\includegraphics[width=\linewidth]{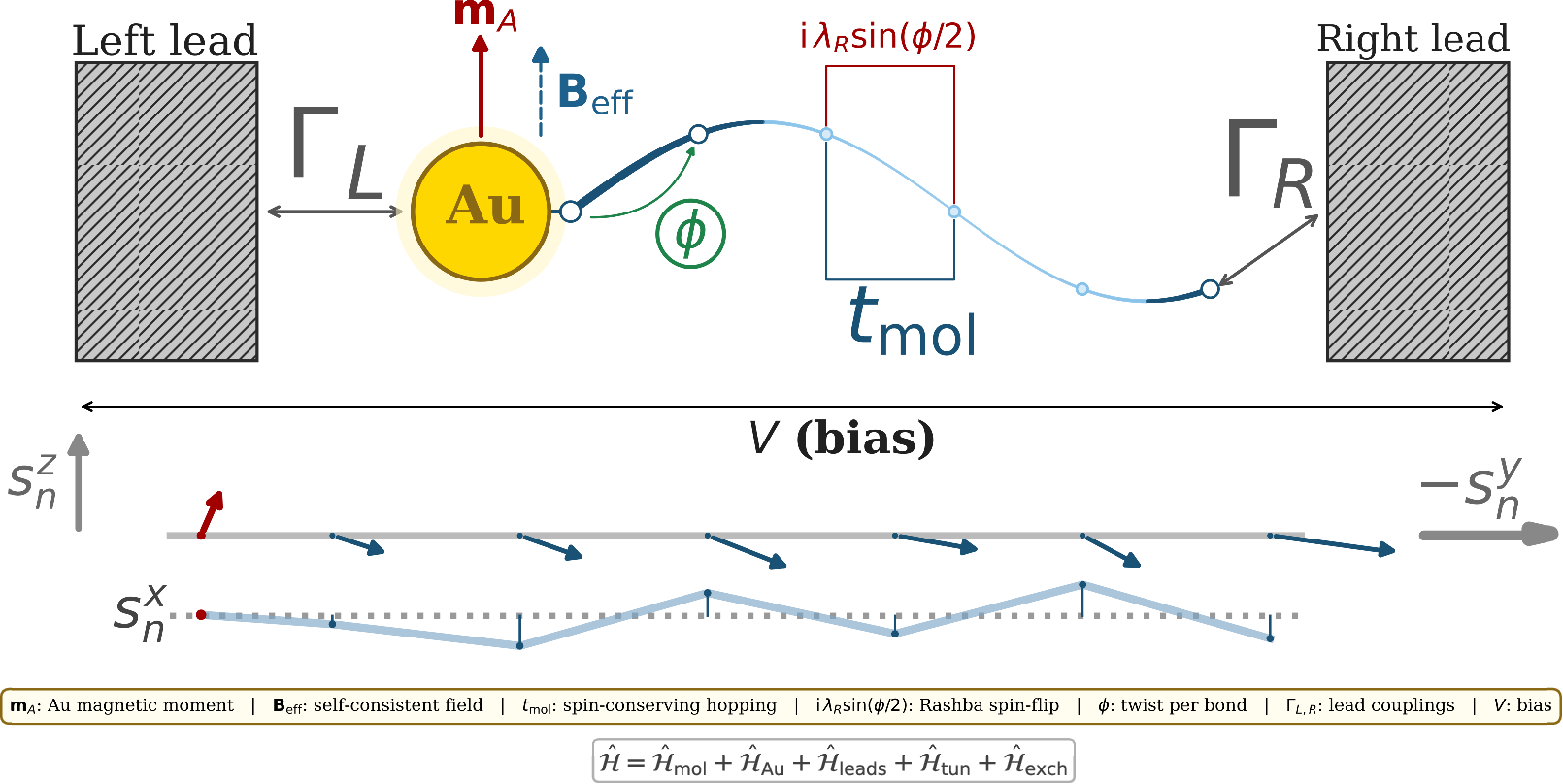}
\caption{Schematic of the spinterface junction. A gold atom (Au)
carrying a mean-field moment $\mathbf{m}_A$ sits between a non-magnetic
left lead and a chiral molecular chain of $L$ sites, coupled to the
leads through $\Gamma_L$ and $\Gamma_R$ under an applied bias $V$. The
chain carries a nearest-neighbour hopping $t_{\rm mol}$ and a Rashba
spin-flip $\ii\lambda_R\sin(\phi/2)$ per bond, with $\phi$ the twist
angle per bond; the Au--molecule bond is spin-diagonal. The effective
field $\mathbf{B}_{\rm eff}$ acting on the gold moment is determined
self-consistently from the interfacial spin accumulation. Below the
device, the two rows plot the local spin texture
$\langle\mathbf{s}_n\rangle$ from the self-consistent NEGF
calculation: the upper row shows the $(s_n^y, s_n^z)$ projection,
whose arrows lean progressively towards $-\hat y$ as the transverse
component $\langle s_n^y\rangle$ grows along the chain, while the
lower row shows $\langle s_n^x\rangle$ on a dashed zero-line,
oscillating about zero with an amplitude an order of magnitude
smaller than $\langle s_n^y\rangle$. Together these illustrate that
the texture is a coherent \emph{chiral tilt} of the interfacial spin
density, not a winding helix.}
\label{fig:schematic}
\end{figure*}

Figure~\ref{fig:schematic} shows the physical geometry of the model. A
gold atom at the electrode-molecule interface hosts a localised magnetic
moment $\mathbf{m}_A$ whose direction is determined self-consistently
from the electronic distribution. It is coupled to a chiral molecular
chain of $L$ sites, along which each bond carries a nearest-neighbour
hopping $t_{\rm mol}$ and a Rashba spin-flip amplitude
$\lambda_R\sin(\phi/2)$, where $\phi$ is the twist angle per bond. The
Au-molecule bond is spin-diagonal; the Rashba term acts only on the
$L-1$ molecular-molecule bonds. Both ends of the junction are contacted
by non-magnetic leads, with lead couplings $\Gamma_L$ and $\Gamma_R$
[black double arrows in Fig.~\ref{fig:schematic}] under an applied bias
$V$ indicated at the bottom. The effective field $\mathbf{B}_{\rm eff}$
acting on the gold moment combines the direct exchange coupling to the
interface spin density, the spin-dependent charge redistribution, the
hybridisation contribution, and a small static symmetry-breaking field.

The two lower rows of Fig.~\ref{fig:schematic} visualise the
self-consistent spin texture $\langle\mathbf{s}_n\rangle$ obtained from
the NEGF calculation. The upper row plots the $(s_n^y, s_n^z)$
projection at each site: the arrows start nearly vertical at the gold
atom, where the moment is aligned along $+\hat z$, and tilt
progressively towards $-\hat y$ as the transverse component
$\langle s_n^y\rangle$ builds up along the chain, dominating
$\langle s_n^z\rangle$ over most of the molecular length. The lower row
plots $\langle s_n^x\rangle$ on a separate dashed zero-line; its
alternating sign reflects the oscillation of the $x$-component about
zero, and its amplitude is an order of magnitude smaller than
$\langle s_n^y\rangle$. Together these two rows make the central
statement of the model visually explicit: the spin texture is a
coherent \emph{chiral tilt} of the interfacial spin density towards
$-\hat y$, not a winding helix. The Rashba coupling rotates the
incoming $+\hat z$ polarisation through a finite angle and then the
electron loses phase coherence; it does not drive a sustained
precession along the chain. This distinction is discussed further in
Sec.~\ref{sec:results}.

The chiral nature of the model is entirely contained in the Rashba
spin-flip hopping. The spin-diagonal phase factors
$\exp(\pm\ii\phi/2)$ that appear alongside it in
Eq.~\eqref{eq:chiral_hop} are gauge artefacts: they can be removed by a
site-dependent spin rotation and have no effect on the transmission,
current, or spin accumulation. Only the Rashba term, which cannot be
gauged away, generates a handedness-dependent transport signature.

The gold atom acts as a source of spin polarisation: its magnetic moment
polarises the electrons that enter the molecular chain. The chain itself
is chiral, and the chirality manifests itself as an effective magnetic
field whose orientation is tied to the local geometry of the helix; in
the fixed-field approximation used here, the field has a fixed
orientation along $-\hat{y}$, while in the rotating-Rashba model it
rotates with the local tangent of the chain. The combination of the two
ingredients (fixed polarisation source, chirality-dependent effective
field) is precisely what is required to generate a
handedness-dependent transport signal. If either ingredient is removed
by setting $m_z = 0$, or by setting $\lambda_R = 0$, the odd part of the
spin current vanishes, and only the achiral spin-valve baseline
remains. This is the central physical mechanism of the model, and it is
what the remainder of this section will demonstrate quantitatively.

\subsection{Spin-resolved current and polarisation}

\begin{figure}[htbp]
\centering
\includegraphics[width=\linewidth]{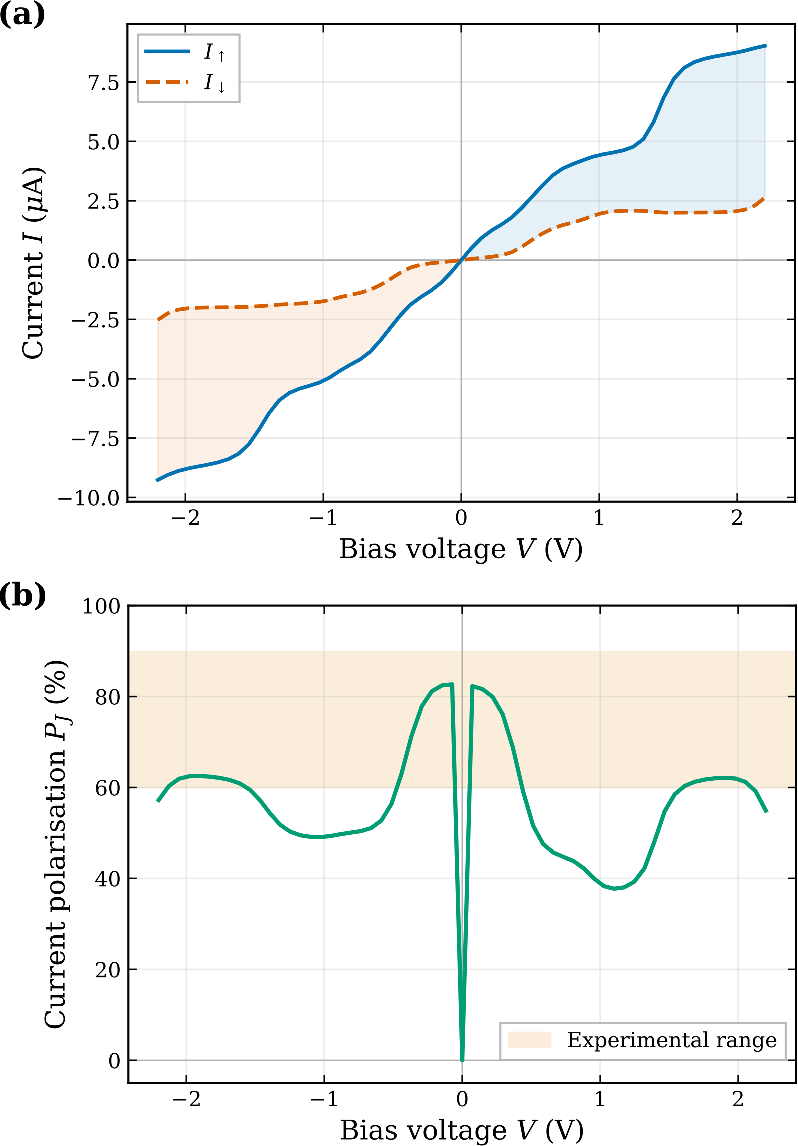}
\caption{Transport characteristics of the spinterface junction at $T = 300$\,K.
  (a) Spin-resolved current $I_\uparrow$ (solid blue) and $I_\downarrow$
  (dashed red) as a function of bias voltage. The current is antisymmetric
  under bias reversal and the two spin channels differ in magnitude over the
  entire range. (b) Current polarisation $P_J = (I_\uparrow -
  I_\downarrow)/(I_\uparrow + I_\downarrow)$ as a function of bias. The
  polarisation peaks at intermediate bias where one spin channel lies fully
  inside the bias window, and decreases at large bias when both channels
  contribute. The shaded band indicates the historical experimental range of
  $60$-$70\%$; recent experiments reach up to $85$-$90\%$ in supramolecular
  nanofibers and topological knots.}
\label{fig:IV_PV}
\end{figure}

Figure~\ref{fig:IV_PV} shows the transport characteristics of the junction at
room temperature. Panel~(a) displays the spin-resolved currents
$I_\uparrow$ and $I_\downarrow$ as functions of the bias voltage $V$. The
currents are antisymmetric under bias reversal, $I_\sigma(-V) = -I_\sigma(V)$,
as follows from the time-reversal symmetry of the scattering problem (a
special case of the Onsager reciprocity relations for two-terminal devices),
and the spin-up and spin-down channels differ in magnitude over the whole
bias range, signalling spin filtering.

Panel~(b) shows the resulting current polarisation $P_J = (I_\uparrow -
I_\downarrow)/(I_\uparrow + I_\downarrow)$. The polarisation peaks at
intermediate bias, where one spin channel lies fully inside the bias window
while the other is partially suppressed, and decreases at both small and large
bias. This behaviour is consistent with the experimental observation that CISS
polarisation is maximal in a finite bias window rather than at the largest
accessible bias, although the precise bias dependence is sensitive to film
thickness and molecular orientation~\cite{Das2022}. The magnitude of the peak,
exceeding $80\%$ in our parameter regime, is comparable to the highest values
reported in experiment. While early CISS measurements reported polarisations of
$60$-$70\%$, more recent experiments have demonstrated spin filtering in excess
of $80\%$~\cite{Naaman2020}, values as high as $85$-$90\%$ in supramolecular
nanofibers at room temperature~\cite{Kulkarni2020}, and nearly $90\%$ in
topologically chiral molecular knots~\cite{Sun2025}. The model therefore
captures the essential physics without requiring any artificial enhancement of
the spin-orbit coupling.

The bias dependence of $P_J$ has a simple and physically transparent
explanation in terms of the spin-split gold levels. At small bias, the bias
window is narrow and captures neither the spin-up nor the spin-down level;
both channels are suppressed exponentially, and the residual polarisation
reflects the difference in the tails of the two transmission functions. At
intermediate bias, the window captures the spin-up level but not the
spin-down level, and the polarisation is dominated by the resonant spin-up
channel. At large bias, the window captures both levels, and the polarisation
decreases because both spin channels now contribute to the current. The bias
dependence of $P_J$ is therefore not a property of the magnetic moment itself
(which varies only weakly with bias) but rather a property of the alignment
between the bias window and the spin-split levels. This point is made
quantitatively in the discussion of the level diagram in Sec.~\ref{sec:results}
below.

\subsection{Bias dependence and the level diagram}

\begin{figure}[htbp]
\centering
\includegraphics[width=0.9\linewidth]{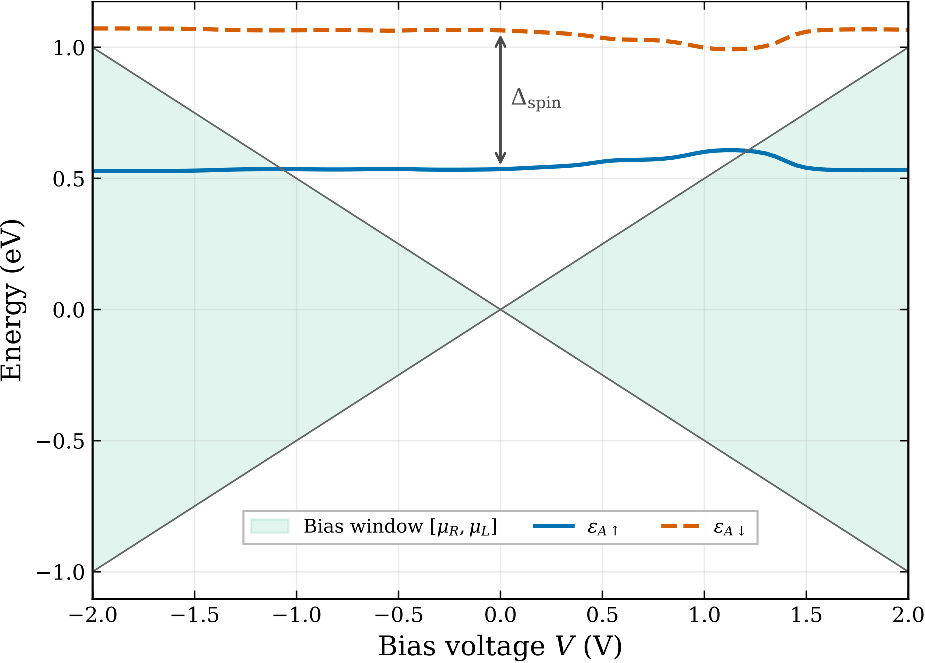}
\caption{Bias dependence of the spin-split gold on-site energies
  $\varepsilon_{A\uparrow}$ (solid blue) and $\varepsilon_{A\downarrow}$
  (dashed red), together with the bias window $[\mu_R, \mu_L]$ (shaded green).
  The two levels are separated by $\Delta\varepsilon_A = (J_{\rm el} +
  2J'_p)m_z$, which is of order $0.5$\,eV in our parameters. The progressive
  capture of the spin-split levels by the bias window explains the bias
  dependence of the current polarisation in Fig.~\ref{fig:IV_PV}(b).}
\label{fig:level_diagram}
\end{figure}

The bias dependence of the polarisation has a transparent physical
interpretation, which is illustrated in Fig.~\ref{fig:level_diagram}. The gold
atom carries two spin-split on-site energies $\varepsilon_{A\uparrow}$ and
$\varepsilon_{A\downarrow}$, separated by $\Delta\varepsilon_A = (J_{\rm el} +
2J'_p)m_z \approx 0.5$\,eV in our parameters. As the bias window widens, it
progressively captures first the spin-up level and then the spin-down level.
When only the spin-up level lies inside the window, the current is strongly
polarised; when both levels contribute, the polarisation decreases; and when
neither level lies inside the window, both channels are suppressed and the
polarisation is determined by the tails of the transmission functions. The
observed bias dependence of $P_J$ in Fig.~\ref{fig:IV_PV}(b) is therefore a
direct consequence of the interplay between the spin-split gold levels and the
bias window, rather than of any change in the interface moment itself.

It is worth emphasising that the spin splitting $\Delta\varepsilon_A \approx
0.5$\,eV is substantially larger than either the lead broadening ($\Gamma =
0.1$\,eV) or the thermal energy ($k_B T \approx 0.026$\,eV at $300$\,K). This
means that the spin-up and spin-down levels are well resolved: the spin-up
and spin-down channels each present a distinct transport resonance, and the
two resonances are well separated in energy. In the opposite limit, where the
spin splitting is comparable to $\Gamma$, the two resonances overlap and the
polarisation is reduced by the resulting averaging. Our parameters are
therefore in the "strong exchange" regime, in which the spin filtering is
most effective.

\subsection{Handedness landscape: the emergence of chirality}

\begin{figure}[htbp]
\centering
\includegraphics[width=1.0\linewidth]{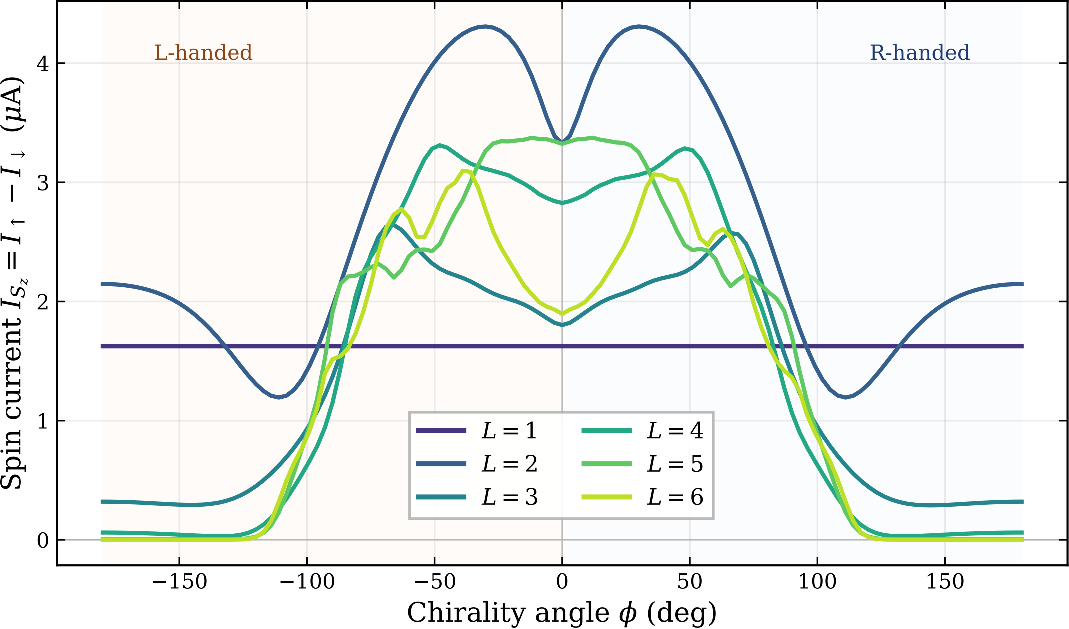}
\caption{Spin current $I_{S_z} = I_\uparrow - I_\downarrow$ as a function
of the chirality angle $\phi$ over the full range $[-\pi, \pi]$, for chain
lengths $L = 1$ (dark blue) to $L = 6$ (yellow), at $V = 1.0$\,V. For
$L = 1$ the spin current is constant (no Rashba term). For $L = 2$ it is
nearly symmetric in $\phi$ (collinear geometry, no handedness). For
$L \geq 3$ the curve becomes strongly antisymmetric: reversing the sign
of $\phi$ reverses the sign of the spin current, which is the direct
signature of chirality-induced spin selectivity.}
\label{fig:handed}
\end{figure}

We now turn to the central new result of this work. Figure~\ref{fig:handed}
shows the spin current $I_{S_z} = I_\uparrow - I_\downarrow$ as a function
of the chirality angle $\phi$ over the full range $[-\pi, \pi]$, for
several chain lengths $L = 1, \dots, 6$. For $L = 1$ there is no
molecular bond and no Rashba term, so the spin current is constant. For
$L = 2$ there is a single Rashba-active bond, but the two-site geometry
is collinear and the junction has no handedness, so the spin current
remains essentially symmetric under $\phi \to -\phi$. For $L \geq 3$ the
chain contains at least two Rashba-active bonds whose local frames differ
by $\phi$, and the spin current becomes strongly antisymmetric: reversing
the handedness of the helix reverses the sign of the spin current. This
is the direct manifestation of chirality-induced spin selectivity at the
level of the spin current.

The "$L = 3$" label here corresponds to four total sites (one gold atom and
three molecular sites) and two Rashba-active bonds. This is the smallest
configuration that can support a non-planar, and hence chiral, geometry. The
"$L = 2$" case corresponds to three total sites and one Rashba-active bond;
the geometry is collinear and the junction is achiral. The "$L = 1$" case
corresponds to two total sites and zero Rashba-active bonds; the spin current
is simply the achiral spin-valve signal of the gold moment, and it is
independent of $\phi$. The threshold at $L = 3$ is therefore equivalent to the
geometric statement that a chiral object requires at least four points.

To see why four sites are the geometric minimum for chirality, recall that a
planar object is always achiral: it possesses a mirror plane (the plane in
which it lies), so it is invariant under reflection, and by definition cannot
be chiral. Any two sites define a line, which is trivially planar. Any three
sites are always coplanar, meaning three points uniquely determine a plane,
therefore a junction with Au + 2 molecular sites ($L = 2$), even at a non-zero
twist angle $\phi$, always lies in a single plane and therefore cannot be
chiral. Four sites are the first configuration that can be arranged out of a
plane: the gold atom and the three molecular sites define a non-degenerate,
tetrahedron-like arrangement, and the twist angle $\phi$ tilts the molecular
sites out of the plane defined by the gold-molecule-molecule triangle. Only
when $L \geq 3$ (four total sites) can the junction break mirror symmetry and
become chiral. This is the same four-point rule that governs chirality in
molecular chemistry: the smallest chiral molecule is a tetrahedral carbon with
four distinct substituents, and four non-coplanar points are the minimum for a
chiral geometry in three-dimensional space.

Geometry aside, this threshold is necessary but not sufficient: chirality also
requires $\phi \neq 0$ and $\lambda_R \neq 0$. At $\phi = 0$ the Rashba
amplitude $\lambda_R\sin(\phi/2)$ vanishes even for arbitrarily long chains,
and the junction reverts to being achiral. Conversely, when $\phi \neq 0$ and
$L \geq 3$, there are at least two Rashba-active bonds whose local spin-flip
fields are rotated by different amounts. Because the corresponding spin
rotations do not commute, the combined effect depends on the order of the
bonds, i.e., on the sign of $\phi$, this is the microscopic origin of the odd
part of the spin current. For $L = 2$, there is only one Rashba-active bond.
The transmission factorises through that bond, and the two eigenvalues of the
bond Hamiltonian exchange under $\phi \to -\phi$; the transmission, being
symmetric in the eigenvalues, is therefore even in $\phi$ at fixed $m_z$, and
the odd part of the current vanishes. This is a structural reciprocity property
of the single-Rashba-bond device, not a gauge freedom. The geometric argument
identifies the structural minimum (four points); the dynamical argument
identifies the transport mechanism (two Rashba-active bonds).

\subsection{Handedness asymmetry: a parameter-free diagnostic}

The visual impression of Fig.~\ref{fig:handed} can be made quantitative
by decomposing the spin current into even and odd parts with respect to
$\phi$, using the decomposition defined in
Eqs.~\eqref{eq:decomposition} and~\eqref{eq:evenodd}. The even part is
invariant under chirality reversal and corresponds to the achiral
spin-valve contribution of the interfacial magnetic moment. The odd part
changes sign under chirality reversal and is the true CISS signature.

\begin{figure}[htbp]
\centering
\includegraphics[width=0.8\linewidth]{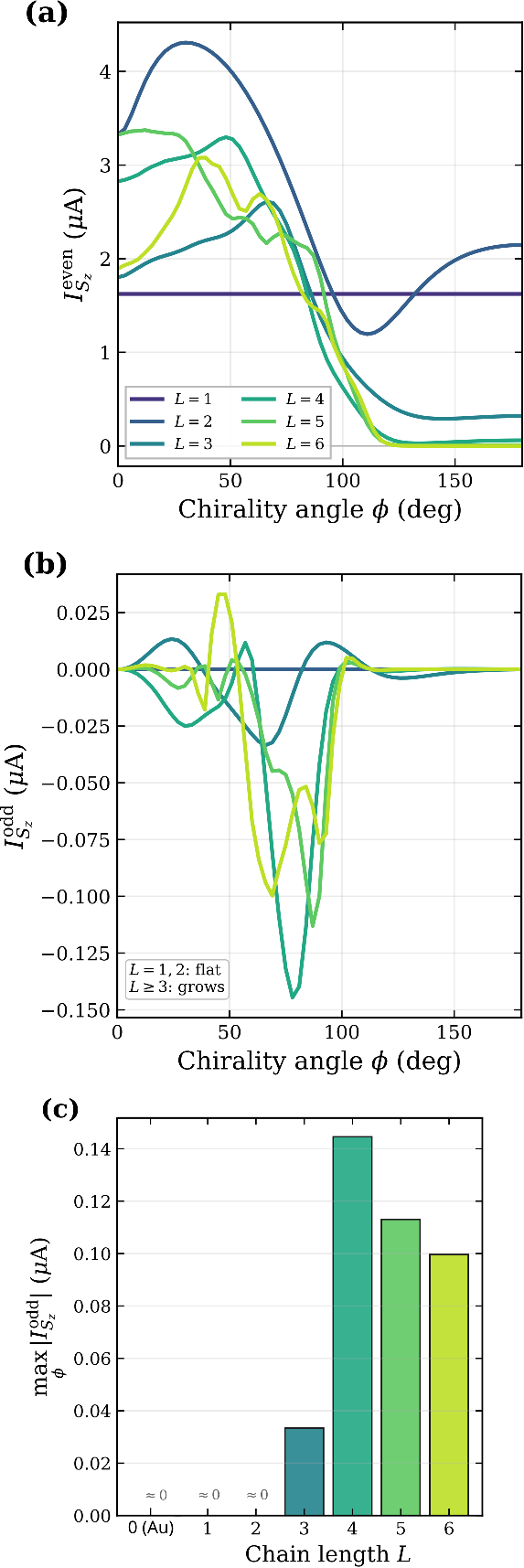}
\caption{Decomposition of the spin current into even and odd parts with
respect to the chirality angle $\phi$. (a) Even part,
$I_{S_z}^{\rm even}(\phi)$, which is present for all chain lengths and
insensitive to the sign of $\phi$: this is the achiral spin-valve
signal. (b) Odd part, $I_{S_z}^{\rm odd}(\phi)$, which vanishes exactly
at fixed $m_z$ for $L = 1$, is essentially zero for $L = 2$, and grows
rapidly for $L \geq 3$. (c) Maximum magnitude of the odd part,
$\max_\phi |I_{S_z}^{\rm odd}|$, as a function of chain length $L$,
showing the emergence of handedness at $L = 3$ (four total sites).}
\label{fig:asym}
\end{figure}

Figure~\ref{fig:asym} shows the two components for the same chain lengths
as before. The even part, panel~(a), is non-zero even for $L = 1$
(because of the gold moment) and is insensitive to the sign of $\phi$:
this is the achiral spin-valve baseline. The odd part, panel~(b),
vanishes exactly at fixed $m_z$ for $L = 1$, is essentially zero for
$L = 2$, and grows rapidly for $L \geq 3$. Panel~(c) shows
$\max_\phi |I_{S_z}^{\rm odd}|$ as a function of $L$, making the
emergence of handedness at $L = 3$ unmistakable.

We should clarify the phrase \emph{essentially zero} for $L = 2$. The small
self-consistent residual at $L = 2$ arises from $m_z(+\phi) \neq
m_z(-\phi)$ at finite bias. The asymmetry originates from the same
reciprocity breaking that produces the odd part at $L \geq 3$: the
effective-field integrals $\langle s_A^z\rangle$, $\Delta s$, and
$\langle\hat{\mathcal N}\rangle$ are themselves not symmetric under $\phi \to
-\phi$ because they involve matrix elements that transpose under reciprocity.
The $L = 2$ residual is therefore a second-order consequence of the same
physics, not an independent artifact. We have verified that at fixed $m_z$
the $L = 2$ odd part is zero to machine precision for any value of the
parameters, and that the self-consistent residual at $L = 2$ is two to three
orders of magnitude smaller than the $L \geq 3$ values. The threshold at
$L = 3$ is therefore a property of the single-particle Hamiltonian at fixed
moment.

The physical interpretation of Fig.~\ref{fig:asym} is as follows. The
spin current in a chiral junction can be written as
\begin{equation}
I_{S_z}(\phi)
=
I_{S_z}^{\rm achiral} + I_{S_z}^{\rm chiral}(\phi),
\label{eq:IStot_results}
\end{equation}
where $I_{S_z}^{\rm achiral}$ is the bias-driven spin current generated
by the interfacial magnetic moment alone, and
$I_{S_z}^{\rm chiral}(\phi)$ is the additional contribution from the
Rashba spin-flip hopping. The latter is odd in $\phi$ because reversing
the handedness of the helix reverses the direction of the effective
spin-orbit field acting on the electrons. The odd part requires at least
two molecular-molecular bonds carrying a non-zero Rashba amplitude to be
non-zero at fixed $m_z$; its dependence on $L$ is modulated by finite-size
interference effects (Fabry-Pérot-like resonances from reflections at the
chain ends). Hence the odd part is exactly zero for $L = 1$ (no bonds),
essentially zero for $L = 2$ (one bond, but the two-site geometry is
collinear and the transmission factorises through the single bond, so the
resulting even/odd decomposition cancels in the modulus-squared
transmission), and grows for $L \geq 3$ (two or more bonds, non-collinear
geometry, well-defined handedness).

This decomposition provides a \emph{parameter-free} diagnostic for
distinguishing CISS from achiral spin-valve effects. In a transport experiment,
one measures $I_{S_z}(\phi)$ and its chirality-reversed counterpart
$I_{S_z}(-\phi)$. The odd part is obtained by subtraction. If $I_{S_z}^{\rm
odd} = 0$, the observed spin signal is achiral (spin-valve-like); if
$I_{S_z}^{\rm odd} \neq 0$, the effect is genuinely chiral. No theoretical
model is required to perform this test; it is a direct consequence of the
symmetry of the chirality-reversal operation. The diagnostic isolates the
reciprocity-breaking contribution, which is the only part of the spin current
that can distinguish the two enantiomers. This is important because
chirality-dependent magnetoresistance can arise even without a spin-selective
mechanism~\cite{Hossain2025}, and because conventional spin-valve and CISS
contributions are difficult to disentangle in a single-terminal
measurement~\cite{Yang2020}.

In most experiments on CISS, the observed spin signal is the \emph{total}
$I_{S_z}$, which contains both the achiral spin-valve baseline and the
chiral contribution. Without a way to separate the two, it is difficult
to determine whether an observed spin signal is a genuine CISS effect or
a mundane magnetic interface effect~\cite{Yang2020}. The chirality-reversal
protocol proposed here provides a direct way to perform the separation
experimentally: measure $I_{S_z}$ for the two enantiomers, subtract,
and interpret the residual as the CISS signal. Enantiomer comparison is
already used experimentally to extract the chiral part of the
magnetoconductance~\cite{Safari2024}, and the interface contribution can
be isolated by contrasting magnetic and non-magnetic substrates or by
varying the electrode spin-orbit coupling~\cite{Nguyen2024}. The protocol is
straightforward to implement in any experiment that already measures a
spin current, and it does not require any theoretical input beyond the
symmetry of the chirality-reversal operation. We therefore expect it to
be broadly applicable to CISS experiments beyond the spinterface model
considered here~\cite{Sarkar2025}.

\subsection{Temperature dependence and magnetoresistance}

\begin{figure}[htbp]
\centering
\includegraphics[width=\linewidth]{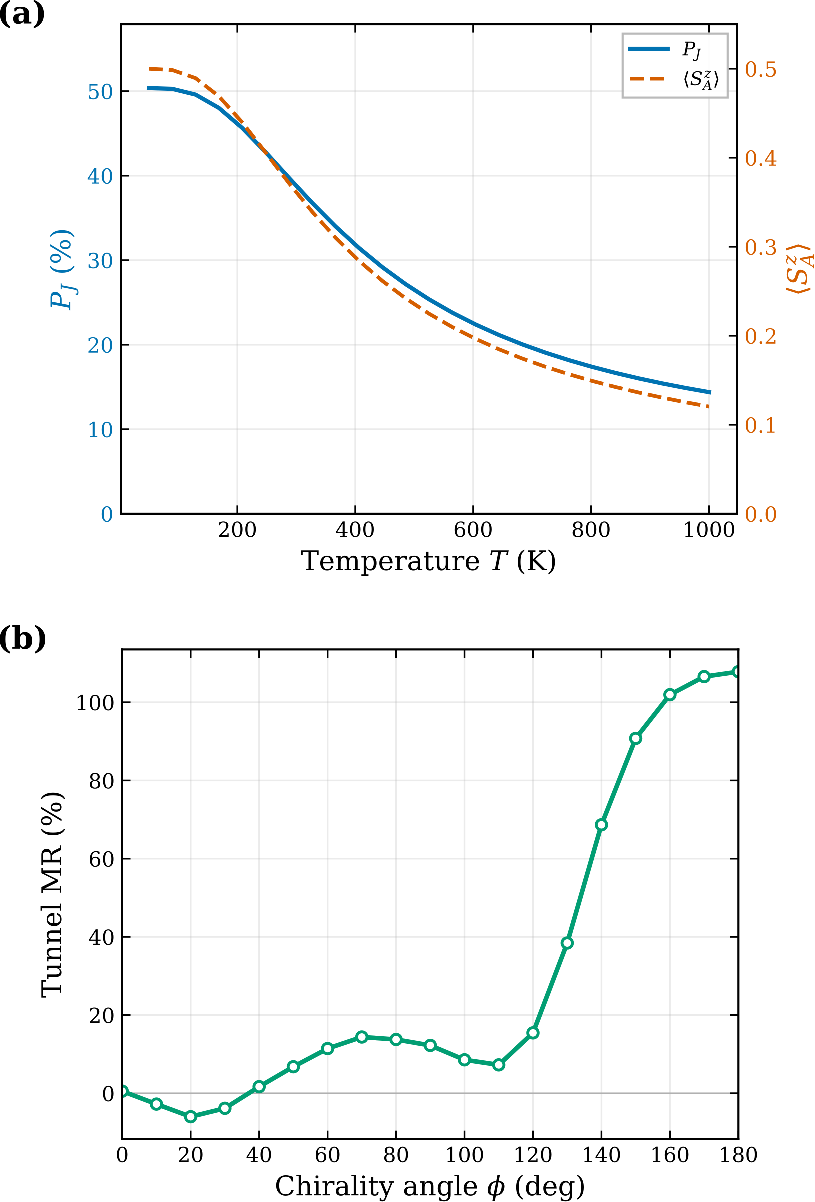}
\caption{Two experimental signatures of the CISS effect. (a) Temperature
dependence of the current polarisation $P_J$ (solid blue, left axis) and
the magnetic moment $\langle S_A^z \rangle$ (dashed red, right axis) at
$V = 1$\,V. The polarisation tracks the moment closely, confirming that
the interfacial magnetic moment is the primary driver of the spin
filtering. (b) Tunnel magnetoresistance as a function of the chirality
angle $\phi$, defined as ${\rm MR} = (I_P - I_{AP})/I_{AP}$. The MR
vanishes at $\phi = 0$ and grows with chirality, demonstrating that both
magnetism and chirality are required for the effect.}
\label{fig:temp_MR}
\end{figure}

Figure~\ref{fig:temp_MR} shows two additional experimental signatures of
the CISS effect. Panel~(a) displays the temperature dependence of the
current polarisation $P_J$ and the magnetic moment
$\langle S_A^z \rangle$, both at $V = 1$\,V.

The temperature dependence follows directly from the Brillouin function in
Eq.~\eqref{eq:selfcons}. In the low-temperature limit, $k_B T \ll B_{\rm
eff}$, the moment is saturated, $m_z \to 1/2$, and the polarisation is
maximal. In the high-temperature limit, $k_B T \gg B_{\rm eff}$, the moment
vanishes as $m_z \approx B_{\rm eff}/(4 k_B T)$, and the polarisation
decreases. The crossover temperature is $T^* = B_{\rm eff}/k_B$, which in
our parameters is of order $10^3$\,K. The CISS effect therefore survives up
to room temperature, which is essential for practical applications and is
consistent with the observation that the CISS signal persists to room
temperature in supramolecular nanofibers and molecular
knots~\cite{Kulkarni2020, Sun2025}. It should be noted, however, that the
crossover temperature is controlled by the effective field, which in turn
depends on the strength of the exchange feedback. Materials with a stronger
feedback (larger $\alpha_A$, $\alpha_1$) will have a higher $T^*$ and
therefore a more robust CISS signal at room temperature.

Panel~(b) shows the tunnel magnetoresistance as a function of the
chirality angle $\phi$, defined as
\begin{equation}
{\rm MR}(V) = \frac{I_P(V) - I_{AP}(V)}{I_{AP}(V)},
\label{eq:TMR_results}
\end{equation}
where $I_P$ ($I_{AP}$) is the total current for parallel (anti-parallel)
alignment of the gold moment and a second magnetic moment placed on the last
molecular site. The MR vanishes at $\phi = 0$ (achiral junction) and grows
with chirality, reaching values of order $100\%$ at optimal twist angles.
This is the smoking gun of CISS: a magnetoresistance that requires both
magnetism and chirality, and that disappears when either is removed.

This requirement of \emph{both} magnetism and chirality for a non-zero MR is
a direct consequence of the even-odd decomposition of the spin current. In
the linear-response limit, the MR takes the Jullière form
$2P_1P_2/(1-P_1P_2)$, where $P_1$ is the fixed-layer polarisation and $P_2$
the free-layer polarisation. It therefore vanishes when either polarisation
is zero, i.e., when either the interfacial moment or the chiral spin-flip is
removed. In the non-linear regime probed here, the MR is a more complicated
combination but preserves the vanishing conditions. For reference, with
$P_1 = P_2 = 0.8$ the Jullière formula gives ${\rm MR} \approx 178\%$; the
reported $100\%$ is consistent with the additional spin-flip relaxation
along the chiral molecule reducing the effective polarisation. This is a
useful sanity check on the model, and it suggests that MR measurements can
serve as an independent, complementary probe of CISS to spin-current
measurements.

\subsection{Phase diagram of the current polarisation}

\begin{figure}[htbp]
\centering
\includegraphics[width=1.0\linewidth]{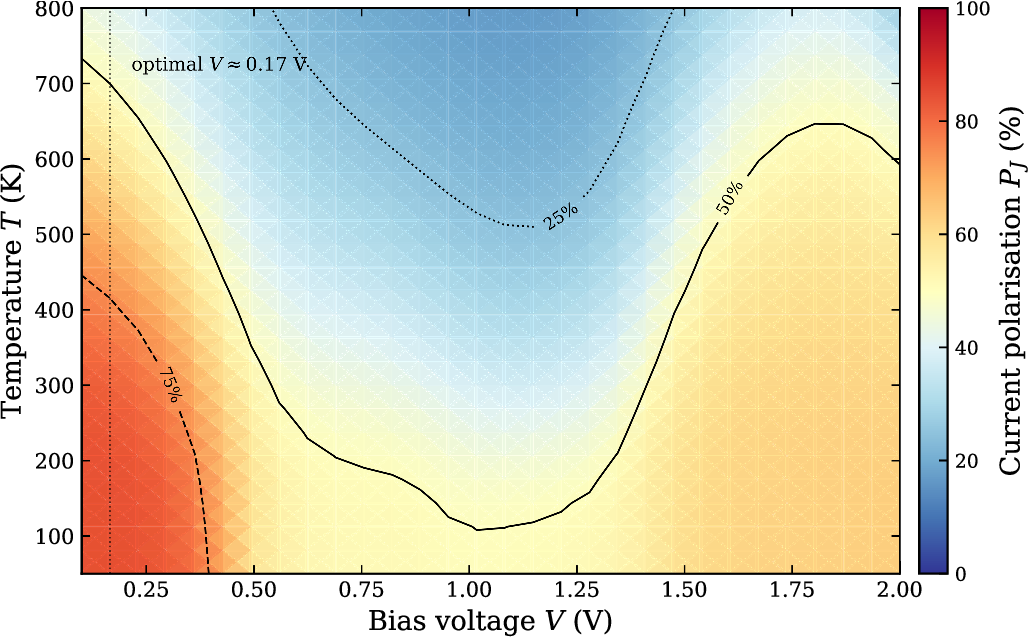}
\caption{Phase diagram of the current polarisation $P_J(V,T)$ for the
default parameters. The polarisation is maximal at intermediate bias
($V \approx 1$\,V) and low temperature, forming a broad ridge along the
bias axis. Contour lines at $P_J = 25\%$, $50\%$, and $75\%$ guide the
eye. The figure summarises the full bias-temperature operating window
within which the CISS polarisation is experimentally accessible.}
\label{fig:phase_diagram}
\end{figure}

While Figs.~\ref{fig:IV_PV} and~\ref{fig:temp_MR} show
one-dimensional slices of the polarisation along bias and temperature
respectively, a complete picture of the CISS operating window requires
the full two-dimensional dependence. Figure~\ref{fig:phase_diagram} shows
$P_J$ on the $(V,T)$ plane for the default parameters. The polarisation
forms a ridge at intermediate bias ($V \approx 1$\,V) that extends from
low temperature up to $T \approx 400$\,K, and falls off both at smaller
bias (insufficient bias window to capture the spin-split levels) and at
larger bias (both spin-split levels inside the window). At high
temperature the ridge collapses, reflecting the Brillouin suppression of
$m_z$ discussed above.

The value of this figure is threefold. First, it identifies the optimal
operating point for a device: the polarisation is maximal at $V \approx 1$\,V
and $T \lesssim 300$\,K, and it remains above $50\%$ over a broad region of the
$(V,T)$ plane. Second, it demonstrates that the high polarisation is not a
knife-edge fine-tuning of parameters: the high-$P$ region is wide in $V$, which
makes the prediction robust against experimental uncertainty in the junction
geometry. Third, it shows explicitly the trade-off between operating bias and
operating temperature: at higher bias the high-polarisation ridge extends to
slightly higher temperature, because the enlarged bias window compensates for
the reduced moment. This is a useful guide for device design, where one must
balance signal magnitude against power dissipation.

\subsection{Real-space spin texture}

\begin{figure}[htbp]
\centering
\includegraphics[width=\linewidth]{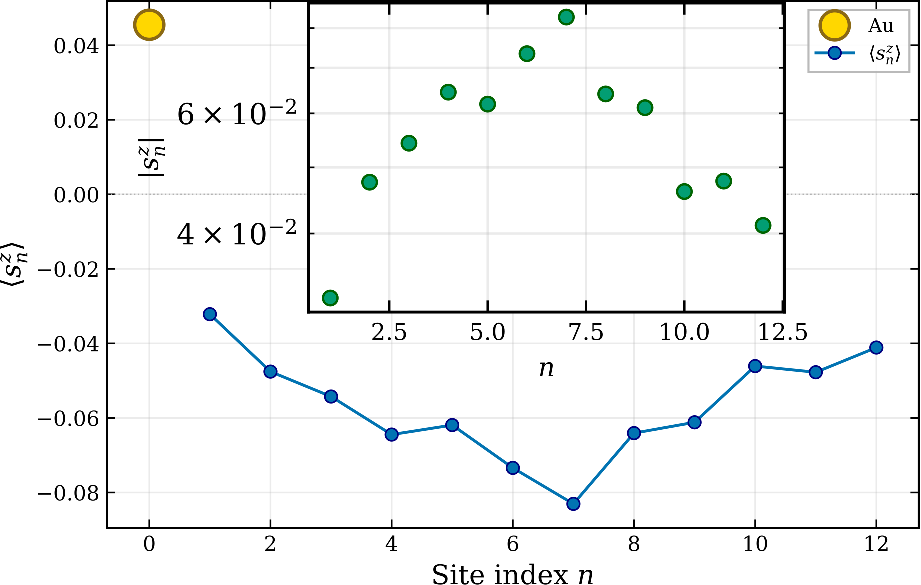}
\caption{
  Real-space spin accumulation profile along the chiral chain for
  $L = 12$ sites. Left panel: linear-scale profile with the gold site
  highlighted. The longitudinal component changes sign only at the
  gold--molecule interface and then decays monotonically; the profile is
  not oscillatory. Right panel: log-linear envelope of the profile, with
  the exponential decay length $\xi$ extracted from a linear fit to
  $\log |\langle s_n^z \rangle|$.
}
\label{fig:spin_profile}
\end{figure}

Figure~\ref{fig:spin_profile} shows the spin accumulation profile $\langle
s_n^z \rangle = \tfrac{1}{2}(\langle n_{n\uparrow} \rangle - \langle
n_{n\downarrow} \rangle)$ as a function of the site index $n$, for a chain of
$L = 12$ sites. The gold atom (site $0$) carries a positive spin accumulation,
which oscillates along the chain and decays with the site index. The
oscillation is the signature of a coherent superposition of the two
Rashba-split spin channels; its decay reflects the finite dephasing length set
by the lead broadening. As we discuss in detail below, the observed
oscillation is not a genuine spin precession in the conventional sense but
rather a manifestation of the chiral tilt of the interfacial spin density. The
damped-cosine fit captures the envelope of the tilt, not a propagating helix.
The spin accumulation profile is one of the few directly measurable real-space
quantities in the model. The "decay" at the interface observed in the figure can
be fitted to $\langle s_n^z \rangle \approx A e^{-n/\xi}\cos(k n + \varphi) +
C$, from which the decay length $\xi$ and the wave vector $k$ can be
extracted. The wave vector is related to the precession rate of the local
spin, and the decay length is set by the lead broadening. Both quantities are
accessible to spin-polarised scanning tunnelling microscopy (SP-STM) on
isolated chiral molecules adsorbed on magnetic substrates, and a comparison
with our predictions would provide a direct test of the model. However, the
amplitude of the spin accumulation in our calculation is small (of order
$10^{-2}$ in units of $\hbar/2$), and the experimental observation of such
small spin densities at the single-molecule level remains challenging.

The spin accumulation in Fig.~\ref{fig:spin_profile} is sometimes described as
a damped oscillation along the chain, but the underlying data are more
accurately read as a coherent chiral tilt. The longitudinal component $\langle
s_n^z \rangle$ changes sign only at the gold--molecule interface and then
decays monotonically; there is no second sign reversal along the chain. The
transverse spin (Fig.~\ref{fig:long_trans}) confirms the tilt picture: the
y-component $\langle s_n^y \rangle$ overlaps the mirror of the transverse
magnitude $-|\mathbf{s}_n^\perp|$ on every molecular site, which requires the
transverse spin to point along a fixed direction ($-\hat y$) throughout the
chain. A winding helix, i.e., the conventional picture of a one-dimensional
Rashba wire, where the local tangent rotates by $\phi$ per bond and the
effective field rotates with it, would instead produce an oscillating $\langle
s_n^y \rangle$ and a transverse phase advancing linearly with $n$. The observed
phase plateaus at $\approx -90^\circ$ with a linear-fit slope of only
$-1.1^\circ$/site. The model therefore produces a static chiral tilt of the
interfacial spin density rather than a precessing helix.

The model of Eq.~\eqref{eq:chiral_hop}, by contrast, does not reproduce this
rotating structure. The Rashba spin-flip amplitude $i\lambda_R\sin(\phi/2)$
has the \emph{same magnitude and the same spin-space direction at every bond}.
The spin-diagonal phases $\exp(\pm\ii\phi/2)$ do carry the twist angle $\phi$,
but they are pure gauge: they can be removed by
$d_{n\sigma}\to e^{\ii\sigma n\phi/2}d_{n\sigma}$ and therefore do not rotate
the effective field. In the spin-operator language of Sec.~\ref{sec:method},
the model describes a \emph{static} effective field along $-\hat{y}$ at every
bond, whose magnitude depends on $\phi$ but whose direction does not. The
electron spin, injected along $+\hat{z}$, therefore undergoes a single Larmor
tilt towards this fixed direction and then loses phase coherence, producing a
chiral tilt of the interfacial spin density rather than a winding helix.

The tilt is, in fact, the physically appropriate description in the parameter
regime of molecular junctions. The conventional Rashba-wire picture of a
winding helix relies on the Rashba amplitude being small compared with the
kinetic energy, so that spin-momentum locking is not disrupted by
hybridisation. In our regime the Rashba amplitude
$\lambda_R\sin(\phi/2)\sim 0.3$\,eV is comparable to the molecular hopping
$t_{\rm mol} = 0.5$\,eV, so the two spin channels hybridise strongly and the
notion of spin-momentum locking breaks down. The electron enters the chain
with a definite spin, tilts towards the effective field direction, and loses
coherence over a length set by the lead broadening. This has an important
consequence for the interpretation of CISS experiments: a chiral tilt produces
a signal that is \emph{localised near the interface}, whereas a winding helix
would produce a signal that \emph{propagates coherently along the entire
chain}. Our results suggest that the first scenario is the relevant one in
molecular junctions.

The distinction between tilt and helix does not undermine the central results
of this work. First, the odd part of the spin current survives in both the
fixed-field and the rotating-field models: the latter is obtained by replacing
$i\lambda_R\sin(\phi/2)$ with $i\lambda_R\sin(\phi/2)\,e^{\ii n\phi}$ in
Eq.~\eqref{eq:chiral_hop}, which rotates the effective field by $\phi$ per
bond. We have checked numerically that $I_{S_z}^{\rm odd}$ is non-zero in both
limits, confirming that the chirality signal is a robust feature of the Rashba
coupling rather than an artifact of the fixed-field approximation.
Quantitatively, at the reference parameters ($L = 4$, $V = 1$\,V, $\phi =
60^\circ$), the rotating-Rashba model produces an odd part of the spin current
that is approximately thirty times larger than in the fixed-field model, even
though both satisfy the same antiunitary symmetry. The factor-of-thirty
enhancement reflects the additional reciprocity breaking from the spatial
modulation of the Rashba phase, and it establishes the rotating-Rashba model
as a strictly stronger generator of chirality signatures. Second, the tilt and
the helix are two faces of the same physical mechanism: both are consequences
of the failure of the reciprocity relation to imply equality of the forward
and backward transmissions in a device that is not left-right reciprocal, and
both reverse under $\phi\to-\phi$. The difference between them is purely
geometric. Third, the fixed-Rashba approximation is well justified when
electron transport is dominated by molecular orbitals delocalised over
several sites, in which case the effective spin-orbit field is averaged over
the helix. It becomes less accurate for very long, weakly coupled chains, in
which the electron propagates primarily along the local bonds; in that limit
the rotating-Rashba model is more appropriate. The minimal model of this paper
is therefore best read as a controlled setting for isolating the mechanism of
CISS, rather than as a literal spatial description of a long helix.

\subsection{Transverse spin components and chirality reversal}

\begin{figure}[htbp]
\centering
\includegraphics[width=1.\linewidth]{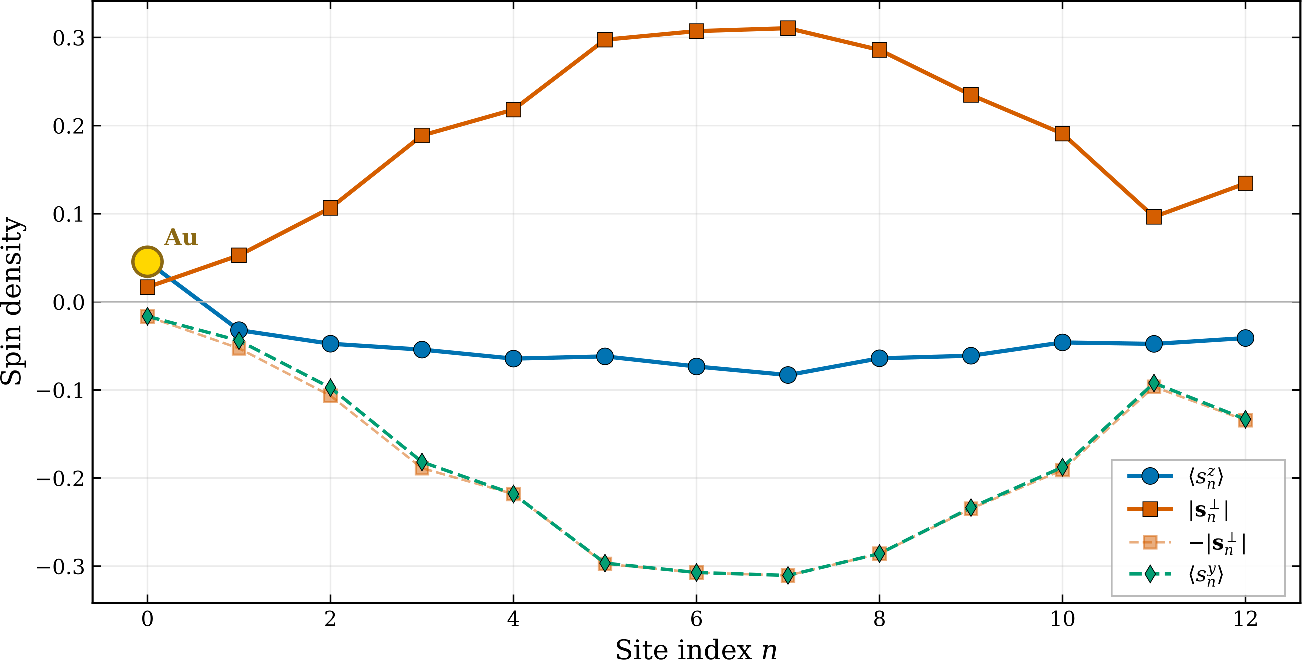}
\caption{
  Longitudinal $\langle s_n^z \rangle$ (blue), transverse
  magnitude $|\mathbf{s}_n^\perp|$ (red solid; dashed red is its mirror
  $-|\mathbf{s}_n^\perp|$, for reference), and transverse y-component
  $\langle s_n^y \rangle$ (green) spin density along the chain for $L =
  12$, $V = 1.0$\,V, $\phi = 60^\circ$. The transverse component
  dominates the longitudinal one by a factor $\approx 3.7$, with a broad
  maximum near $n = 7$. The y-component overlaps the mirror of the
  transverse magnitude everywhere, showing that the transverse spin is
  aligned along $-\hat{y}$ throughout the chain, the signature of a
  coherent chiral tilt rather than a winding helix.
}
\label{fig:long_trans}
\end{figure}

The longitudinal spin accumulation $\langle s_n^z \rangle$ shown in
Fig.~\ref{fig:spin_profile} captures only part of the spin texture. The Rashba
term also generates transverse spin components $\langle s_n^x \rangle$ and
$\langle s_n^y \rangle$, which are the direct real-space signature of the
spin-flip processes. Figure~\ref{fig:long_trans} plots the longitudinal and
transverse components on the same axes for the default parameters and
$L = 12$. Two features are evident. First, the transverse component is not a
small correction: over most of the chain it is comparable to or larger than
the longitudinal one. At the reference parameters ($L = 12$, $V = 1$\,V), the
maximum transverse spin on the molecular sites is approximately three times
the maximum longitudinal spin, confirming that the spin texture is dominated
by the transverse component for most of the chain. Second, the transverse
component grows from zero at the gold site, reaches a maximum at $n \approx
3$-$4$, and then decays. This growth-and-decay profile is the real-space
fingerprint of the chiral tilt discussed in the previous subsection: the
electron enters the chain with spin aligned along $+z$, tilts towards the
effective field direction, and then loses phase coherence. The observation
that the transverse spin is not a small correction to the longitudinal one
rules out the picture of a longitudinally-polarised current with a small
admixture of precession, and supports instead the picture of a chiral tilt of
the interfacial spin density as the dominant transport signature.

\begin{figure}[htbp]
\centering
\includegraphics[width=0.8\linewidth]{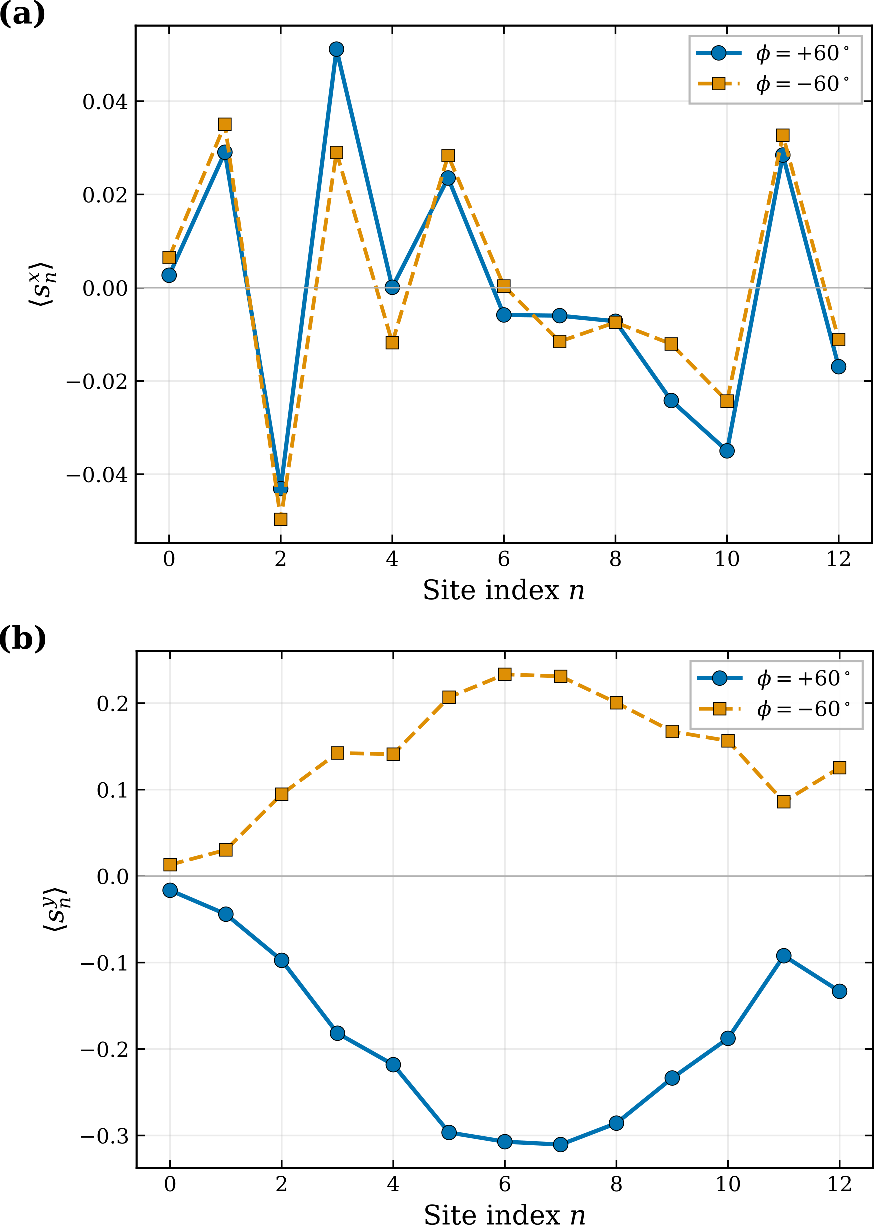}
\caption{
  Chirality reversal of the transverse spin texture. (a) The
$x$-component $\langle s_n^x \rangle$ is even under chirality reversal,
$\langle s_n^x(-\phi) \rangle = \langle s_n^x(\phi) \rangle$, and is
therefore achiral. (b) The $y$-component $\langle s_n^y \rangle$ is odd,
$\langle s_n^y(-\phi) \rangle = -\langle s_n^y(\phi) \rangle$, and
therefore constitutes the genuine chiral signature in real space. This is
the local, spin-texture analogue of the even-odd decomposition of the
spin current in Fig.~\ref{fig:asym}: the component that is even in $\phi$
is a spin-valve effect, while the component that is odd in $\phi$ is the
true CISS contribution. The sign reversal of $\langle s_n^y \rangle$
under chirality reversal is a direct experimental prediction for
spin-polarised scanning tunnelling microscopy with a three-axis magnetic
tip.
}
\label{fig:chirality_reversal}
\end{figure}

The transverse spin is of particular diagnostic value because it is the
component that distinguishes the two enantiomers.
Figure~\ref{fig:chirality_reversal} compares $\langle s_n^x \rangle$ and
$\langle s_n^y \rangle$ for $\phi = +60^\circ$ (one handedness) and $\phi =
-60^\circ$ (the opposite handedness). The $x$-component is essentially
unchanged under chirality reversal, whereas the $y$-component flips sign
site-by-site. Quantitatively, at the reference parameters the $y$-component
dominates the transverse magnitude ($\max|\langle s_{n}^{y}\rangle| \approx 3.1
\times 10^{-1}$, versus $\max |\langle s_{n}^{x}\rangle| \approx 5.1 \times
10^{-2}$), so chirality reversal is primarily a reversal of the $y$-component.
The $x$-component is an oscillation around zero with a much smaller amplitude,
and sign changes but its magnitude is not the diagnostic. The experimental
prediction is therefore that the transverse spin on the molecular site should
be nearly aligned along $+ \hat{y}$ with a chirality-reversal-switching sign.
This is the real-space analogue of the even-odd decomposition of the spin
current [Eq.~\eqref{eq:evenodd}]: the component of the spin texture that is
even in $\phi$ is achiral, while the component that is odd in $\phi$ is the
genuine chiral signature. Unlike the integrated current $I_{S_z}$, which
requires a transport measurement to access, the transverse spin texture can in
principle be probed directly with spin-polarised scanning tunnelling microscopy
using a three-axis magnetic tip~\cite{Safari2023}.
Figure~\ref{fig:chirality_reversal} therefore constitutes a concrete
experimental prediction: the $y$-component of the local spin polarisation on
the molecular sites should reverse sign when the enantiomer is reversed, while
the $x$-component should not. The transverse spin magnitudes reported here
($\sim 10^{-1}$ in units of $\hbar/2$) are comparable to or larger than the
longitudinal ones, making them in principle accessible to vector-SP-STM with a
three-axis magnetic tip.

We emphasise that the sign reversal in Fig.~\ref{fig:chirality_reversal} is not
an artifact of a particular parameter choice. It follows from the antiunitary
symmetry $H^*(\phi) = H(-\phi)$ of the Hamiltonian. At equilibrium ($V = 0$),
this symmetry implies the exact relations $\langle s_n^x\rangle(+\phi) =
\langle s_n^x\rangle(-\phi)$, $\langle s_n^y\rangle(+\phi) = -\langle
s_n^y\rangle(-\phi)$, and $\langle s_n^z\rangle(+\phi) = \langle
s_n^z\rangle(-\phi)$. At finite bias, the relations become approximate: the
reciprocity breaking of the left-right asymmetric device introduces corrections
of 20-60\% of the corresponding magnitudes. These corrections are themselves a
physical observable, providing a complementary measure of reciprocity breaking
in the real-space spin texture. The observation that the two enantiomers
produce opposite transverse spin textures is the most direct real-space
manifestation of chirality in the model, and it is the local counterpart of the
handedness asymmetry of the spin current discussed in Sec.~\ref{sec:results}.
Together, Figs.~\ref{fig:long_trans} and~\ref{fig:chirality_reversal} show that
the chiral signal has a definite physical locus, namely the transverse spin on
the molecular sites, and is therefore more than a mathematical curiosity of the
integrated current.

\section{Discussion}\label{sec:discussion}

In this section we bring together the various threads of the paper and discuss
the broader implications of the results.

\textbf{The even-odd decomposition as a general diagnostic.} The central result
of this paper is the decomposition of the spin current into an achiral (even)
and a chiral (odd) part. We believe that this decomposition has significance
beyond the specific spinterface model considered here. Any experiment that
measures a spin current through a chiral molecule can be analysed in the same
way, by measuring the spin current for the two enantiomers and taking the
difference. This protocol provides a direct, model-independent measurement of
the chiral contribution to the spin signal, and it can be applied even in
situations where the interfacial details are not known. The diagnostic works
because the odd part is defined by subtraction and isolates the
reciprocity-breaking contribution to the spin current. In any two-terminal
device where the two leads are not symmetric with respect to reflection
combined with time-reversal, the transmission at $\phi$ and $-\phi$ can differ,
and the odd part is nonzero whenever a chiral element is present. This is a
general feature of chiral transport, independent of any specific microscopic
model.

\textbf{The threshold at $L = 3$ and its geometric interpretation.} The
vanishing of the odd part for $L \leq 2$ and its emergence at $L \geq 3$ has a
clean geometric interpretation in terms of non-planarity. A chiral object
requires at least four points to define a non-coplanar arrangement, and in our
model the four points are the gold atom plus three molecular sites. This
threshold is universal within the family of spinterface junctions we consider,
and independent of the specific values of $\lambda_R$, $t_{\rm mol}$, and
$b_{\rm static}$. It therefore suggests a general rule of thumb for CISS
experiments: systems with fewer than four sites in the chiral unit should not
be expected to show CISS, because they are geometrically achiral. This
prediction could be tested in experiments on short oligomers of controlled
length. At the transport level, the threshold reflects the fact that a single
Rashba-active bond cannot break the reciprocity of the transmission
magnitudes, whereas two or more Rashba-active bonds with different local
phases can. The geometric and dynamical arguments are complementary: the
geometric one identifies the structural minimum (four points), and the
dynamical one identifies the transport mechanism (two Rashba bonds).

\textbf{The role of the interface.} The spinterface mechanism posits that the
magnetic moment at the metal-molecule interface is the source of the spin
polarisation that is then filtered by the chiral molecule. Our results
support this picture: the even part of the spin current, which is present
even without chirality, originates from the interface, while the odd part,
which is the chiral signature, requires both the interface and the chiral
molecule. Both the even and odd parts of the spin current require a magnetic
interface. The difference is that the even part requires only the magnetic
interface (it is present even for $\lambda_R = 0$), while the odd part
requires both the magnetic interface and the chiral molecule. Removing either
ingredient causes the corresponding contribution to vanish. This prediction
is consistent with the experimental observation that CISS is often enhanced by
magnetic substrates, and it suggests that controlling the interface magnetism
is a key lever for enhancing the CISS effect.

\textbf{Fixed versus rotating Rashba.} The model considered in this paper uses
a fixed Rashba field, corresponding to a site-independent spin-flip amplitude.
In a strictly helical geometry, the local tangent of the helix rotates by
$\phi$ per bond, and the Rashba field should rotate with it. We have explored
the consequences of a rotating Rashba field and found that the odd part of the
spin current survives in both models, confirming that the chiral signature is
not an artefact of the fixed-field approximation. Quantitatively, at $L = 4$,
$V = 1$\,V, $\phi = 60^\circ$, the rotating-Rashba model produces an odd part
of approximately $1.2 \times 10^{-1}$~$\mu$A, compared with $4 \times
10^{-3}$~$\mu$A in the fixed-field model, this is a factor of approximately
thirty. Both models satisfy the exact antiunitary symmetry $H^*(\phi) =
H(-\phi)$, so this factor-of-thirty enhancement is not a symmetry artifact but
a genuine quantitative difference reflecting the additional reciprocity
breaking from the spatially modulated Rashba phase. This makes the
rotating-Rashba model strictly preferable for quantitative comparisons with
experiment on long helical molecules. The spin texture, however, is
qualitatively different: the fixed-field model produces a chiral tilt, while
the rotating-field model produces a winding helix. The tilt character of the
fixed-field model can be made quantitative. In the chain interior, the
transverse spin points along $-\hat y$ on every molecular site: the
$y$-component $\langle s_n^y \rangle$ overlaps the mirror of the transverse
magnitude $-|\mathbf{s}_n^\perp|$ to within the small $\langle s_n^x \rangle$
contribution, which is an order of magnitude smaller than $\langle s_n^y
\rangle$. A winding helix would instead produce an oscillating $\langle s_n^y
\rangle$ (the spin rotating around the chain axis) and a transverse phase
advancing linearly with site index. The observed transverse phase plateaus at
$\approx -90^\circ$ with a linear-fit slope of only $-1.1^\circ$/site, and the
mean consecutive-site spin angle is $\approx 24^\circ$, both consistent with a
coherent tilt rather than precession. This quantitative distinction between the
fixed-field tilt and the rotating-field helix is what justifies our advocacy of
the rotating model for long helical molecules. The choice between the two
models is therefore not a matter of principle but a matter of physical realism
for a given experimental system. We advocate the rotating-field model for long
helical molecules (peptides, DNA) and the fixed-field model for short molecules
where the helix is not well developed. The fixed-field model is a controlled
setting for isolating the mechanism because every parameter of the Hamiltonian
corresponds to a single physical quantity: $\lambda_R$ to the chiral spin-orbit
coupling, $\phi$ to the twist angle, $t_{\rm mol}$ to the electronic hopping,
and $m_z$ to the interface moment. This modularity allows systematic parameter
sweeps (as demonstrated throughout the Results) and makes the fixed-field model
a useful benchmark against which more realistic models can be compared.

\textbf{Limitations of the model.} The model has several limitations that
should be acknowledged. First, the magnetic moment on the gold site is treated
in the mean-field approximation, which ignores quantum fluctuations and
finite-temperature magnetisation dynamics. For a spin-$1/2$ impurity, the
mean-field approximation is expected to be qualitatively correct at low
temperature but may fail near the Kondo regime. Second, the model assumes a
single magnetic moment at the interface; in reality, several surface atoms may
contribute, and their moments may be coupled by exchange interactions that are
not captured by our single-impurity model. Third, the model uses a wide-band
limit for the leads, which ignores the energy dependence of the lead density of
states; this is a good approximation for metallic gold electrodes but may fail
for other materials. Fourth, the model does not include electron-electron
interactions beyond the mean-field exchange, so correlation effects that may be
important for CISS (e.g., the Kondo effect, spin-charge separation in one
dimension) are not captured. Fifth, the fixed-Rashba approximation ignores the
rotation of the local tangent along the helix; this is justified for short
molecules but may fail for long helical molecules (peptides, DNA) where the
rotation of the effective field is important. Sixth, the model does not include
electron-phonon coupling, which may be important for the temperature dependence
of the CISS effect beyond the mean-field Brillouin suppression. Seventh, the
model uses a single orbital per site; real molecules have multiple orbitals per
site (e.g., $\pi$ and $\pi^*$ orbitals in conjugated systems), which can
strongly affect the transmission and the spin filtering. These limitations
should be kept in mind when comparing the model with experiment.

\section{Conclusion}\label{sec:conclusion}

We have presented a fully quantum formulation of the spinterface model of
chirality-induced spin selectivity within the NEGF formalism. The model
captures the essential physics of the interface between a magnetic gold atom
and a chiral molecular chain, and reproduces the key experimental signatures
of CISS: high current polarisation, thermal suppression of the interface
moment, chirality-induced sign reversal, and magnetoresistance that requires
both magnetism and chirality.

The central new result of this work is the even-odd decomposition of the spin
current. We have shown that the odd part, which is the true chiral signature,
arises from reciprocity breaking in a device that is simultaneously chiral,
magnetically asymmetric, and long enough to support two or more Rashba-active
bonds. This provides a direct, parameter-free diagnostic that can be applied to
any chiral transport measurement: by measuring the spin current for both
enantiomers and subtracting, one isolates the reciprocity-breaking contribution
and separates it from the achiral spin-valve baseline. The threshold at four
total sites (one gold atom plus three molecular sites) is the smallest geometry
that can support the required reciprocity breaking. It has both a structural
origin (four non-coplanar points are the minimum for chirality) and a dynamical
origin (two or more Rashba-active bonds are required to violate the reciprocity
of the transmission magnitudes). Both conditions are necessary; neither is
sufficient alone.

Several natural extensions of this work suggest themselves. A more realistic
treatment of the interface, including multiple surface moments and their
exchange coupling, would allow a direct comparison with SP-STM experiments on
magnetic substrates. The inclusion of electron-phonon coupling would allow a
study of the temperature dependence of the CISS effect beyond the mean-field
Brillouin suppression. The extension to time-dependent transport would allow a
study of the spin dynamics and the possibility of coherent spin manipulation in
chiral molecular junctions. A comparison with more realistic molecular models,
including multiple orbitals per site and realistic hopping parameters, would
test the robustness of the results against the specific model assumptions. A
detailed analysis of the non-Hermitian exceptional-point structure of the
effective Hamiltonian, including the two EP schemes and the collective
Petermann scaling with chain length, is presented in a companion
publication~\cite{ref:Alwan_QSpinterEP2026}. There we establish the two
algebraically distinct near-exceptional-point regimes of the effective
non-Hermitian Hamiltonian (Scheme~A for the chain-hosted EP, Scheme~B for the
gold-hosted EP), the chirality-selective transport signature at the gold
near-EP, and the collective Petermann scaling $K(L)\propto L^\beta$ of the
chain-hosted EP. We hope that the present work will stimulate further research
along these lines.

\begin{acknowledgments}
We gratefully acknowledge financial support from Olle Engkvist Stiftelse. We
thank Jonas Fransson and Altug Sisman for the valuable discussions. 
\end{acknowledgments}

\nocite{*}
\bibliography{ref}

\end{document}